\documentclass[sigconf]{acmart}

\usepackage{amscd,amsfonts,amsbsy,rotating}
\usepackage{balance}
\usepackage{graphicx}
\usepackage{epsfig,epstopdf}
\usepackage{subfigure}
\usepackage{multirow}
\usepackage{booktabs}
\usepackage{color,xcolor}
\usepackage{url}
\usepackage{latexsym,bm}
\usepackage{enumitem,balance,mathtools}
\usepackage{wrapfig}
\usepackage{euscript}
\usepackage{algorithm}
\usepackage{algorithmic}
\usepackage{ifpdf}
\usepackage{diagbox}
\usepackage{caption}
\usepackage{makecell}
\usepackage{bm}
\usepackage{sidecap}

\usepackage{array}
\newcolumntype{N}{@{}m{0pt}@{}}

\usepackage{lipsum}
\newcommand\blfootnote[1]{%
  \begingroup
  \renewcommand\thefootnote{}\footnote{#1}%
  \addtocounter{footnote}{-1}%
  \endgroup
}

\theoremstyle{plain}

\newtheorem*{definition*}{Definition}

\newcommand{\minisection}[1]{\vspace{5pt}\noindent\textbf{#1.}}

\AtBeginDocument{%
  \providecommand\BibTeX{{%
    \normalfont B\kern-0.5em{\scshape i\kern-0.25em b}\kern-0.8em\TeX}}}

\acmSubmissionID{fp0042}

\copyrightyear{2020}
\acmYear{2020}
\setcopyright{acmcopyright}\acmConference[SIGIR '20]{Proceedings of the 43rd International ACM SIGIR Conference on Research and Development in Information Retrieval}{July 25--30, 2020}{Virtual Event, China}
\acmBooktitle{Proceedings of the 43rd International ACM SIGIR Conference on Research and Development in Information Retrieval (SIGIR '20), July 25--30, 2020, Virtual Event, China}
\acmPrice{15.00}
\acmDOI{10.1145/3397271.3401073}
\acmISBN{978-1-4503-8016-4/20/07}

\begin{document}
\fancyhead{}
\title{A Deep Recurrent Survival Model for Unbiased Ranking}

\author{Jiarui Jin$^{1}$, Yuchen Fang$^1$, Weinan Zhang$^1$, Kan Ren$^2$, Guorui Zhou$^3$, Jian Xu$^3$, Yong Yu$^1$,\\ Jun Wang$^4$, Xiaoqiang Zhu$^3$, Kun Gai$^3$}
% \authornote{This work was done when the first author worked as internship in DiDi AI Labs.}
\affiliation{$^1$Shanghai Jiao Tong University, $^2$Microsoft Research, $^3$Alibaba Group, $^4$University College London}
\email{{jinjiarui97, arthur\_fyc, wnzhang, yyu}@sjtu.edu.cn, kan.ren@microsoft.com, jun.wang@cs.ucl.ac.uk, {guorui.xgr, xiyu.xj, xiaoqiang.zxq, jingshi.gk}@alibaba-inc.com}

\renewcommand{\shortauthors}{J. Jin, et al.}

\begin{abstract}
	Position bias is a critical problem in information retrieval when dealing with implicit yet biased user feedback data. 
	%Unbiased learning-to-rank has sparked great interest for learning true relevance between a given query and a set of documents, from implicit.
	%Recent advances in unbiased learning-to-rank have sparked great interest in alleviating bias issue, especially position bias in the user logs, and training high-performance unbiased ranker with click data.
	Unbiased ranking methods typically rely on causality models and debias the user feedback through inverse propensity weighting.
	%Most existing approaches mainly focus on conducting debiasing of click data and training a ranker based on inverse propensity weighting.
	While practical, these methods still suffer from two major problems.
	First, when inferring a user click, the impact of the contextual information, such as documents that have been examined, is often ignored.
	% On the other hand, only the document-aware information (\emph{e.g.}, clicked document) is considered but the user's rich behavior patterns in the full ranking list are missed.
	Second, only the position bias is considered but other issues resulted from user browsing behaviors are overlooked.
	% they only learn debiased algorithm from the user clicked queries, while do not consider the non-clicked queries which may provide additional information. 
	In this paper, we propose an end-to-end Deep Recurrent Survival Ranking (DRSR), a unified framework to jointly model user's various behaviors, to 
	% (i) jointly model user's various behaviors (\emph{e.g.}, click and observe) in a unified framework; and
	(i) consider the rich contextual information in the ranking list; and
	(ii) address the hidden issues underlying user behaviors, \emph{i.e.}, to mine observe pattern in queries without any click (non-click queries), and to model tracking logs which cannot truly reflect the user browsing intents (untrusted observation).
	%or even may not been aware of.
	% (ii) estimate unbiased ranking scories directly from the truth relevance ground truth over both clicked and non-clicked queries.
	Specifically, we adopt a recurrent neural network to model the contextual information and estimates the conditional likelihood of user feedback at each position.
	We then incorporate survival analysis techniques with the probability chain rule to mathematically recover the unbiased joint probability of one user's various behaviors.
	%To this end, we first formulate the problem as to estimate the probability distribution of conditional click rate for each document at different positions.
	%Second, we propose a deep recurrent model to better capture the user browsing patterns and mathematically derive the joint probability of user behavior.
	%Also, we capture latent information in non-clicked logs by regarding these logs as censored clicked data, where censorship occurs in click information.
	DRSR can be easily incorporated with both point-wise and pair-wise learning objectives.
	%Third, we design a novel objective function covering both click and non-click data, adaptive on either point-wise or pair-wise objective setting.
	The extensive experiments over two large-scale industrial datasets demonstrate the significant performance gains of our model comparing with the state-of-the-arts.
	%Finally, we extensively evaluate our method on both Yahoo search engine and Alibaba recommender system datasets, obtaining favorable results compared with state-of-the-art methods.
	%To the best of our knowledge, it is the first work providing an adaptive perspective for user behavior modeling and contextual unbiased learning which is the core challenge in the learning-to-rank task.
\vspace{-2mm}
\end{abstract}



\begin{CCSXML}
	<ccs2012>
	<concept>
	<concept_id>10002951.10003317.10003338.10003343</concept_id>
	<concept_desc>Information systems~Learning to rank</concept_desc>
	<concept_significance>500</concept_significance>
	</concept>
	</ccs2012>
\end{CCSXML}

\ccsdesc[500]{Information systems~Learning to rank}
\vspace{-2mm}


\keywords{Unbiased Learning-to-Rank; Position Bias; Cascade Model}

\settopmatter{printacmref=false, printfolios=false}


\maketitle

\vspace{-3mm}
{\fontsize{8pt}{8pt} \selectfont
	\textbf{ACM Reference Format:}\\
	Jiarui Jin, Yuchen Fang, Weinan Zhang, Kan Ren, Guorui Zhou, Jian Xu, Yong Yu, Jun Wang, Xiaoqiang Zhu, Kun Gai. In \textit{
		Proceedings of the 43rd International ACM SIGIR Conference on Research and Development in Information Retrieval (SIGIR '20), July 25--30, 2020, Virtual Event, China.} ACM, New York, NY, USA, 10 pages.
	https://doi.org/10.1145/3397271.3401073}

\vspace{-3mm}
\section{Introduction} 
\begin{figure}[t]
	\centering
	\includegraphics[width=0.4\textwidth]{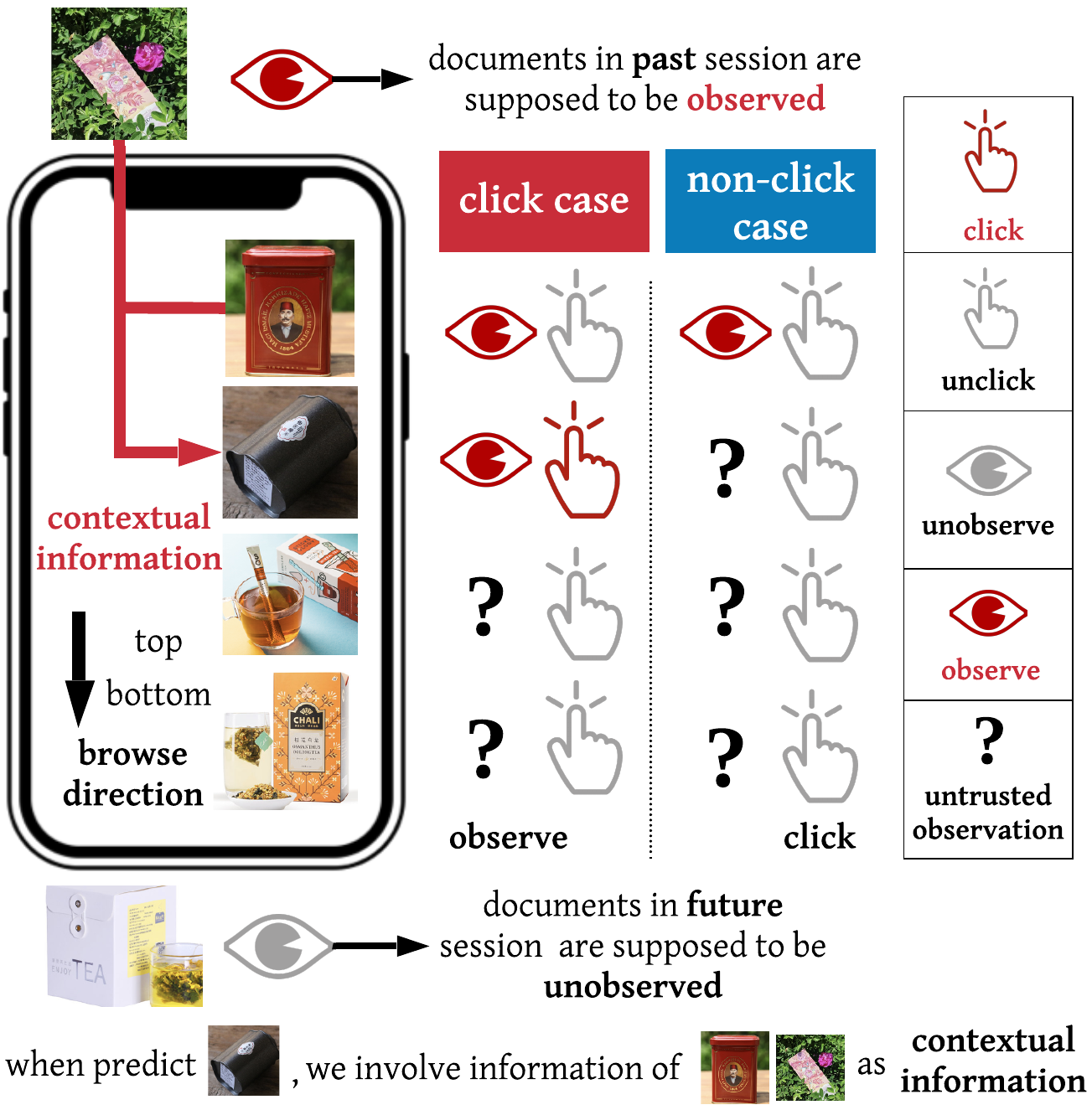}
	\vspace{-3mm}
	\caption {Illustration of user various behaviors (\emph{i.e.}, click and non-click case) when browsing document list as shown in the left side. Notations are provided in the right side.}
	\label{fig:intro}
	\vspace{-7mm}
\end{figure}

% learning to rank in information system
Nowadays, information systems have become a core part for the personalized online services, such as search engines and recommender systems, where machine learning is the key technique for the success \cite{ai2018unbiasedb}.
Among them, learning-to-rank \cite{liu2009learning} is a fundamental approach which learns to present a ranked list of items as the most relevant to the user query or most likely preferred by the target user.
However, there is no ground truth except expensive labeled data from human experts for training a ranker, which facilitates the usage of implicit user feedbacks \cite{joachims2005accurately,wang2016learning}, \emph{i.e.}, user clicks \cite{joachims2017unbiased,wang2016learning} and browsing logs \cite{wang2016learning,ai2018unbiaseda}.
Although the usage of the implicit user feedback alleviates data labeling cost, it introduces the data bias problem \cite{joachims2017unbiased,ai2018unbiaseda}.
As shown in Figure~\ref{fig:intro}, taking position bias as an example, a user typically observes the presented item list from top to down, in such a way the attention of the user drops rapidly and the user may observe and click more likely on the top presented items than the bottom ones \cite{joachims2005accurately}.
Simply optimization for the ranking performance based on the implicit feedback data may result the ranking function in learning the presenting order, rather than the true relevance or the real user preferences.
To tackle such a bias issue, many researchers have explored the potential technical approaches in training a high efficient model with unbiased learning-to-rank.\par

One line of the research is based on counterfactual learning \cite{agarwal2019general,jagerman2019model}, which treats the click bias as the counterfactual factor \cite{rosenbaum1983central} and debiases the user feedbacks through inverse propensity weighting (IPW) \cite{joachims2017unbiased,wang2016learning}. 
% These approaches successfully conduct IPW principle in learning an unbiased ranker from click data and prove that the objective function is an unbiased estimate of the risk function defined on relevance measure.
For instance, \citet{ai2018unbiaseda} and \citet{hu2019unbiased} respectively proposed to employ the dual learning method for jointly estimating position bias and training a ranker. % via point-wise and pair-wise algorithms.
However, these methods either focus on the position of the item while ignoring the \textbf{contextual information} of the given ranking list, \emph{e.g.}, the content of the previous items may influence the observation of the next item \cite{ai2018unbiaseda,hu2019unbiased}, or optimize the ranking performance directly while neglecting the nature of the user browsing behaviors, \emph{e.g.}, the click always happens at the observed item \cite{joachims2017unbiased,wang2016learning}.
Another line of research investigates user browsing behavior model \cite{chapelle2009dynamic,dupret2008user,wang2015incorporating,wang2013content}, where several basic assumptions about the user browsing behaviors are adopted to maximize the likelihood of the observations. 
% in the history data collected from the user browsing logs
For example, \citet{fang2018intervention} extended position-based model and proposed an estimator based on invention harvesting, which, however, only considers the query-level contents (\emph{e.g.}, query feature) rather than the document-level contextual information (\emph{e.g.}, the previous observed documents).
% Moreover, as is discussed in \cite{joachims2017unbiased}, the user browsing models mainly optimize the likelihood maximization through user browsing assumptions, which treats the issue of learning to rank as an afterthought.\par

More importantly, the prior works often focus on addressing the gap between user behavior and true preferences, while leaving latent issues hidden in the user behaviors, to be unsolved:
(i) As Figure~\ref{fig:intro} illustrated, when the user starts a search with a query, she may stop browsing by interruption or end the search session due to lack of interest, which leaves many queries without any user click behavior, often referred as \textbf{non-click queries}.
It is impractical to mine click patterns in these non-click queries.
However, these queries contain large amount of observe patterns, \emph{e.g.}, users are often impatient and only observe several top documents when selecting daily items such as fruits; while becoming patient and browse several sessions before click when selecting luxury items such as phones.
Different from recent investigations on abandoned queries, defined as Web search queries without hyperlink clicks, we here mainly focus on mining observe patterns instead of distinguishing bad and good abandonment \cite{li2009good,song2014context}.
(ii) When the user scrolls down the page and observes the presented items, the system may track (through check points) that the user has observed until the last position of the screen.
However, this may not be true since the user would have stopped and lost her attention before checking the items on the last positions.
It is not realistic to set eyetracking for each user during each of her visit \cite{joachims2005accurately}.
Hence, the tracking logs of the user browsing history cannot truly tell that the users are actually checking the contents, and we called these noisy logs \textbf{untrusted observations}, as Figure~\ref{fig:intro} shows.\par

Based on the above analysis, when designing unbiased learning-to-rank algorithm, the current state-of-the-art methods have not well solved, even may not been aware of, the following challenges, which we address in this paper:
\textbf{(C1)} The user behaviors contain various and highly correlated patterns based on the  \textbf{contextual information}.
\textbf{(C2)} There are large scale of latent observe patterns hidden in the \textbf{non-click queries}.
\textbf{(C3)} The \textbf{untrusted observation}, another unsolved issue, is caused by limitation of tracking logs.

To tackle these challenges, we propose a novel framework called \emph{deep recurrent survival ranking} (DRSR) to formulate the unbiased learning-to-rank task as to estimate the probability distribution of user's conditional click rate.
To capture user behavior pattern, we combine survival model and recurrent neural network (RNN) in DRSR framework. Specifically, the RNN architecture incorporates all the top-down contents in the ranking list as contextual information, while the survival model derives the joint probability of user behavior via the probability chain rule, which enables modeling both observed and censored user behavior patterns \textbf{(C1)}. 
%	Also, we mathematically derive the joint probability of user behavior via the probability chain rule, which enables modeling various behavior patterns in a unified framework \textbf{(C1)}.
We then assume that a user's favored documents in the non-click queries could hide in unobserved ones out of browsing scope.
This is similar to those patients who leave the hospital and die out of investigation period.
Hence, we can leverage survival analysis techniques \cite{cox1972regression,kaplan1958nonparametric,ren2019deep} via 
% regarding that the click information is missed in those non-click logs due to the censored user behavior, and 
treating non-click logs as censored data of clicked ones where the censorship occurs in the click behavior \textbf{(C2)}.
In seeking a proper way to measure relevance for untrusted observation, we model conditional probability and design a novel objective function to learn relative relevance between trusted and untrusted feedbacks in pair-wise setting \textbf{(C3)}.

% extend survival analysis techniques \cite{ren2019deep} to learning-to-rank via regarding \emph{click} as true event.
% Specially, we model that the user ends browsing and \emph{leave}s the result page, which is analogous to that a patent stops investigation and \emph{leave}s hospital observation. 
% By this way, non-clicked logs are treated as the censored data of clicked logs where the censorship occurs in the click information.\par

The major contributions of this paper can be outlined as follows.

% contribution
% 1. content-based unbiased learning, more precise for user propensity modeling
% 2. considering untrusted observations, which, to some extent, corrects the propensity modeling in a novel way.
\begin{itemize}[topsep = 3pt,leftmargin =10pt]
	\item We propose an innovative framework to jointly capture the correlation of user behaviors and train an unbiased rank with contextual information of the rank list.
	\item We incorporate cascade model with survival analysis to deeply mine hidden user observe patterns in non-click queries.
	\item We provide a Pairwsie Debiasing training scheme to model relative relevance between trusted and untrusted observations.
	% \item We propose an innovative framework to jointly capture the corelation of user behaviors and train an unbiased rank with contextual information of the rank list.
	% propose a novel framework to jointly estimate position bias and train a ranker with contextual information of the rank list. 
	% cascading model to better capture the user browsing patterns and mathematically derive the conditional probability of user behaviors, so as to debias the implicit user feedbacks for learning to rank in both point-wise and pair-wise setting.
	% \item We provide two kinds of debias loss functions: Pointwise Debiasing and Pairwise Debiasing, adaptive to both point-wise and pair-wise settings.  
	% we incorporate both click and non-click information and utilize the context-based model for comprehensive user behavior modeling.
\end{itemize}

Extensive experiments on Yahoo search engine and Alibaba recommender system datasets demonstrate the superiority of DRSR over state-of-the-arts. 
To the best of our knowledge, in the unbiased learning-to-rank task, it is the first work providing adaptive user behavior modeling using contextual information with survival analysis.

% The rest of the paper will be organized as follow.
% We will discuss the related works in Section~\ref{sec:related-work}.
% We introduce the general framework for both point-wise and pair-wise unbiased learning-to-rank, and further formulate the problem in Section~\ref{sec:framework}.
% Section~\ref{sec:method} presents our proposed model in detail including components in architecture and formulation in loss function.
% In Section~\ref{sec:exp}, we discuss the experimental setup, report the corresponding results, and provide comprehensive analysis.
% We conclude our paper and show the possible future works in Section~\ref{sec:conclusion}.

\begin{table*}[t]
	\centering
	\caption{A summary of notations in this paper.}
	\vspace{-3mm}
	\begin{tabular}{|l|l|}
		%% \hline
		%% \multicolumn{2}{|c|}{A summary of notations in regular unbiased learning-to-rank section.}\\
		\hline
		$q$, $D_q$, $C_q$, $O_q$, $\mathcal{D}$ & Query $q$ and set of documents $D_q$ associated with click information $C_q$ and observe information $O_q$. \\
		& Dataset $\mathcal{D}$ for all query, formulated as $\mathcal{D} = \{ (q, C_q, O_q, D_q) \}$.\\
		\hline
		$i$, $d_i$, $\bm{x}_i$, $r_i$, $c_i$, $o_i$ & Position $i$ and $i$-th document $d_i$ in $D_q$ with feature vector $\bm{x}_i$, relevance information $r_i$, click information $c_i$ \\
		& and observe information $o_i$. Note that we utilize binary value here and can obtain $o_i = 1$ from $c_i = 1$.\\
		\hline
		% $i$, $I$ & set of $L$ time slices: $l = 1, 2, 3, ..., L$, $l$-th browsing interval $V_l$  defined as $V_i = (t_{l-1}, t_l]$.\\
		$p_i$, $h_i$, $W(i)$, $S(i)$ & At $i$-th document $d_i$, \emph{click probability} (P.D.F) $p_i$ and \emph{relevance probability} (conditional probability) $h_i$ as\\
		& defined in Eqs.~(\ref{eqn:discreteP}).  and (\ref{eqn:discreteH}).
		\emph{observe probability} (C.D.F.) $S(i)$ and \emph{unobserve probability} (C.D.F.) $W(i)$ as\\
		& defined in Eqs.~(\ref{eqn:discreteS}) and (\ref{eqn:discreteW}).\\
		\hline
		$D_q^+$, $D_q^-$, $D_q^*$, $I_q$ & Real positive document set $D_q^+$ 
		% where user has expressed her feedback, i.e. observed and clicked; negative & 
		document set $D_q^-$
		% where user has seen but not given her feedback, i.e. observed and unclicked, 
		and uncertain document set $D_q^*$ 
		% where user feedback is unclear due to censorship, i.e. untrusted. 
		the whole document set \\
		& $D_q$ formulated as $D_q = D_q^+ \bigcup D_q^- \bigcup D_q^*$, $I_q$ denotes the set of document pairs ($d_i$, $d_j$).
		where $d_i$, $d_j$ are \\
		& sampled from two sets of $D_q^+$, $D_q^-$, $D_q^*$.\\
		\hline
	\end{tabular}
	\vspace{-3mm}
	\label{tb:pre}
\end{table*}

\section{Related Work}
\label{sec:related-work}
% In this section, we introduce related work on unbiased learning-to-rank and untrusted observation modeling.
\minisection{Unbiased Learning to Rank}
Learning to rank \cite{liu2009learning} is a fundamental technique for information systems, such as search engine,
recommender system 
and sponsored search advertising. 
There are two streams of unbiased learning to rank methodologies.
One school is based on some basic assumptions about the user browsing behaviors \cite{chapelle2009dynamic,dupret2008user,wang2015incorporating,wang2013content}.
These models maximize the likelihood of the observations in the history data collected from the user browsing logs.
Recently, \citet{fang2018intervention} extended position-based model and proposed an effective estimator based on invention harvesting.
As is discussed in \cite{joachims2017unbiased}, these model only model user behavior patterns without sufficient optimization for learning to rank problem.
The other school derived from counterfactual learning \cite{joachims2017unbiased,wang2018position} which treats the click bias as the counterfactual factor \cite{rosenbaum1983central} and debias the user feedback through inverse propensity weighting \cite{wang2016learning}.
Recently, \citet{ai2018unbiaseda} and \citet{hu2019unbiased} respectively proposed to employ the dual learning method for jointly estimating position bias and training a ranker.
However, these prior works often ignore the rich contextual information in query and omit user's various behaviors except click. 
In this paper,  we propose an innovative approach 
a novel cascade model adaptive in both point-wise and pair-wise setting.
In addition to taking joint consideration of click and non-click data via survival analysis, we also model the whole ranking list through recurrent neural network. 

\minisection{Survival Analysis}
In the field of learning-to-rank, deep learning-based sequential algorithms have won much attention and surpassed several traditional models via considering the time information of each interaction \cite{fang2019deep}.
However, the event occurrence information may be missed, due to the limitation of the observation period or track procedure \cite{wang2019machine}, which is called censorship. 
A key technique addressing censorship is to estimate the probability of the event occurrence at each time typically studied by survival analysis. 
There are two main streams of survival analysis.
The first view is based on traditional statistics scattering in three categories. 
Non-parametric methods \cite{kaplan1958nonparametric,andersen2012statistical}
%% including Kaplan-Meier estimator \cite{kaplan1958nonparametric} and Nelson-Aalen estimator \cite{andersen2012statistical} 
are solely based on counting statistics.
%% , which is too coarse-grained to perform personalized modeling. 
Semi-parametric methods \cite{cox1972regression,tibshirani1997lasso}
%% such as Cox proportional hazard model \cite{cox1972regression} and its variants Lasso-Cox \cite{tibshirani1997lasso} 
assume some base distribution functions with the scaling coefficients.
%% for fine-tuning the final survival rate prediction. 
Parametric models \cite{lee2003statistical} assume that the survival time or its logarithm result follows a particular theoretical distribution.
These methods either base on statistical counting information or pre-assume distributional forms for the survival rate function, which generalizes not very well in real-world situations.
The second school of survival analysis takes from machine learning perspective. 
%% Survival random forest which was first proposed in \cite{gordon1985tree} derives from standard decision tree by modeling the censored data while its idea is mainly based on counting-based statistics. 
These machine learning methodologies include survival random forest \cite{gordon1985tree}, Bayesian models \cite{ranganath2016deep}, support vector machine \cite{khan2008support} and multi-task learning solutions \cite{li2016multi,alaa2017deep}. 
Recently, \citet{ren2019deepa,ren2019deepb} proposed a recurrent neural network model which captures the sequential dependency patterns between neighboring time slices and estimates the survival rate through the probability chain rule.
In this paper, we extend the methodology of the survival analysis to provide fine-grained unbiased ranking list in a unified learning objective without making any distributional assumptions.
% Note that our model takes end-to-end learning in a unified learning objective without making any distributional assumptions, which is capable of addressing censorship and providing fine-grained unbiased ranking list.

\section{preliminary}
\label{sec:preliminary}
% In this section, we extend the untrusted observation into unbiased learning-to-rank both the point-wise and pair-wise settings.
% We then give the formulation of problem to be solved in this paper. 

\subsection{Point-wise Unbiased Learning-to-Rank}
\label{sec:pointwise}
%% In learning-to-rank scenario, ranker $f$ is asked to propose a relevance score $r$ after receiving item information (e.g., position, content, etc.).
%% And in unbiased learning-to-rank scenario, the goal of the ranker is to propose an appropriate a relevance score and return unbiased ranked item list \cite{joachims2005accurately}.\par
The fundamental task in learning-to-rank scenarios is to learn a ranker $f$ which assigns a score $r$ to the document $d$ according to item feature $\bm{x}$.
Then, the documents with respect to the query $q$ return the list in descending order of their scores. 
Let $q$ denotes the query and $D_q$ the set of documents associated with $q$.
We consider three subsets contained in the set of documents $D_q$ as follows:
\begin{itemize}[topsep = 3pt,leftmargin =5pt]
	\item 	$D_q^+$: set of the real positive documents that user has expressed her feedback on, \emph{i.e.}, observed and clicked. 
	\item	$D_q^-$: set of the real negative documents that user has seen but not given her feedback on, \emph{i.e.}, observed and unclicked.
	\item	$D_q^*$: set of the uncertain documents that user feedback is unclear due to user's leave behavior, \emph{i.e.}, untrusted.
\end{itemize} 

We describe $d_i$ the $i$-th document in $D_q$ and $\bm{x}_i$ the feature vector of $d_i$.
Let $r_i$ represent the relevance of $d_i$.
For simplicity we only consider binary relevance here, \emph{i.e.}, $r_i=1$, $r_i=0$ and $r_i = ?$, where $? \in \{0, 1\}$ but unknown for those untrusted documents; and one can easily extend it to the multi-level relevance case.
In the point-wise setting, the risk function in learning is defined on a single data point $x$ as
% \vspace{-0.5mm}
\begin{equation}
R(f) = \int_{q} \int_{d_i \in D_q^+} L(f(\bm{x}_i),r_i) \ d P(\bm{x}_i, r_i) ,
\end{equation}
where $f$ denotes a ranker, $L(f(\bm{x}_i),r_i)$ denotes a point-wise loss function and $P(\bm{x}_i,r_i)$ denotes the probability distribution on $\bm{x}_i$ and $r_i$.
The goal of learning-to-rank is to find the optimal ranker $f$ that minimizes the loss function.
% Suppose that there is a labeled dataset in which the relevance of documents with respect to queries is given. 
% One can learn a ranker $\hat{f}$ through the minimization of the empirical risk function (objective function) as follows:
% \begin{equation}
% \label{eqn:pointwiseranker}
% \hat{f} = \text{arg} \min_{f} \sum_{q}  \sum_{d_i \in D_q^+} L(f(x_i),r_i)
% \end{equation}
% Most ranking measures in information retrieval (IR) \cite{joachims2017unbiased} only utilize relevant documents in their definitions, and thus the loss function here is defined on relevant documents with label $r_i$.
Traditionally, the ranker is learned with labeled data containing user browsing logs.
%nOne can also consider using click data as relevance feedbacks from users, more specifically, viewing clicked documents as relevant documents and unclicked documents as irrelevant documents, and training a ranker with a click dataset.
However, click data is indicative of individual users' relevance judgments, but is also noisy and biased \cite{joachims2017unbiased,ai2018unbiaseda}.
This is what we call biased learning-to-rank.

\begin{figure*}[t]
	\centering
	\includegraphics[width=0.7\textwidth]{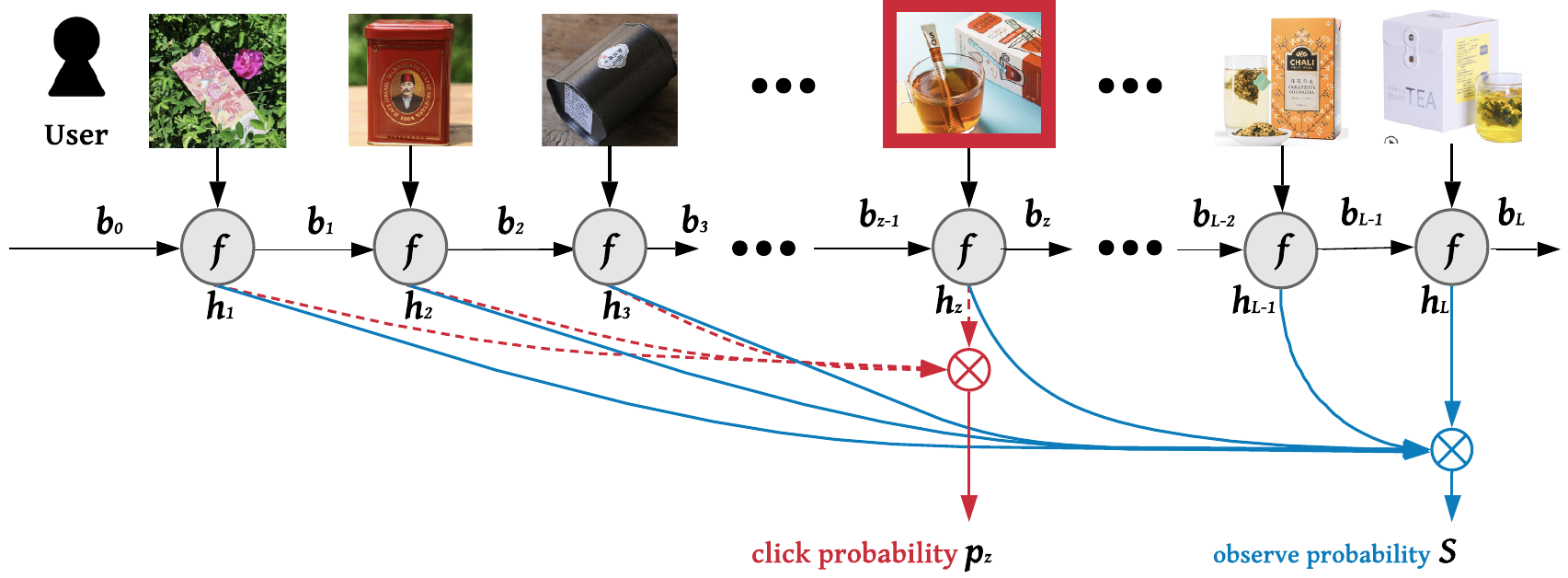}
	\vspace{-3mm}
	\caption {Illustration of Deep Recurrent Survival Ranking model. Note that we mine click patterns in click case and observe patterns in both click and non-click cases.
	}
	\label{fig:ranker}
	\vspace{-3mm}
\end{figure*}

\subsection{Pair-wise Unbiased Learning-to-Rank}
\label{sec:pairwise}
Traditionally, in the pairwise setting, the ranker $f$ is still defined on a query document pair ($\bm{x}$, $r$), and the loss function is defined on two data points: positive document $d_i$ and negative document $d_j$. 
We here also take those untrusted documents into consideration.
% To achieve this, we extend the elements in pair ($d_i$, $d_j$) here.
Specially, we sample $d_i$ and $d_j$ 
% from two different subsets 
from three candidate sets, \emph{i.e.}, $D_q^+$, $D_q^-$ and $D_q^*$ instead of only from the first two sets.
%% $d_i \in D_q^+$, $d_j \in D_q^-$ and $d_k \in D_q^*$.
% We adopt notations in Section~\ref{sec:pointwise}.
%% A detailed illustration is given in Figure~\ref{fig:pairwise}.
Let $q$ denote a query. 
Let $\bm{x}_i$ and $\bm{x}_j$ denote the feature vectors from $d_i$ and $d_j$ respectively. 
Let $r_i$ and $r_j$ represent the document $d_i$ and document $d_j$ respectively.
Let $I_q$ denote the set of document pairs ($d_i$, $d_j$).
Similarly, for simplicity we only consider binary relevance here.
%% and one can easily extend it to the multi-level relevance case. 
The risk function 
% and the minimization of empirical risk function 
is defined as
\begin{equation}
\label{eqn:pairwiserisk}
R(f) = \int_q \int_{(d_i, d_j) \in I_q} L(f(\bm{x}_i),r_i,f(\bm{x}_j),r_j) \ dP(\bm{x}_i,r_i,\bm{x}_j,r_j) ,
\end{equation}
% \begin{equation}
% \label{eqn:pairwiseranker}
% \hat{f} = \text{arg} \min_{f} \sum_{q} \sum_{(d_i, d_j) \in I_q} L(f(x_i),r_i,f(x_j),r_j)
% \end{equation}
where $L(f(\bm{x}_i),r_i,f(\bm{x}_j),r_j)$ denotes a pair-wise loss function.
% $\mathcal{T}_q$ represents pair set containing document pairs, i.e., $(d_i, d_j)$.\par
% One can consider using click data to directly train a ranker, that is, to conduct `biased learning-to-rank' \cite{joachims2005accurately}.\par
% Different with previous pair-wise unbiased learning-to-rank approaches \cite{hu2019unbiased}, we implement conditional possibility to model pair-wise loss function.
%% We introduce our deep unbiased learning-to-rank model in the following section.

\subsection{Problem Formulation}
The key question in any unbiased learning-to-rank algorithm is how to fill the gap between click and relevance.
Given the fact that users click a search document $d_i$ ($c_i = 1$) only when it is both observed ($o_i = 1$) and perceived as relevant ($r_i = 1$), most recent works \cite{joachims2017unbiased,wang2016learning,wang2018position,ai2018unbiaseda} formulate the problem as
$P(c_i = 1 | \bm{x}) = P(c_i = 1 | o_i =1; \bm{x}) \cdot P(o_ i = 1 | \bm{x})$
, where we can define relevance as
\begin{equation}
\label{eqn:relevance}
P(r_ i = 1 | \bm{x}) \doteq P(c_i = 1 | o_i =1; \bm{x}) = \frac{P(c_i = 1 | \bm{x})}{P(o_i = 1 | \bm{x})}.
\end{equation}
The task of biased learning-to-rank is to estimate click and return a ranking list according to $P(c_i = 1 | \bm{x})$, while the aim of unbiased learning to rank is to derive relevance from click data and provide a ranking list according to $P(r_ i = 1 | \bm{x})$.\par

Many click models \cite{dupret2008user,chapelle2009dynamic,guo2009efficient} have investigated how to model the impact from previous clicked document.
To simplify, we only study the session with single click here.
One can easily extend into multiple click session via truncating multiple one into several single ones.
Actually, this sequence truncation method, over the sequential data with multiple events, has been widely used in many works covering various fields such as recommender system \cite{jing2017neural}, conversion attribution \cite{ren2018learning} and survival analysis \cite{ren2019deepa}, which truncates the raw sequences according to the events of interest (\emph{i.e.}, click in our case).\par
% Note that in this case, \emph{click} behavior caused by either user behavior or truncation method indicates the end of browse logs.

\section{Methodology}
\label{sec:method}
% In this part, we present Deep Recurrent Learning-to-Rank (DRL2R) as a method of jointly estimating position bias and training a ranker with both clicked and non-clicked data.
% Furthermore, we apply DRL2R in both point-wise and pair-wise setting.

% \subsection{Preliminaries}
% \label{sec:pre}

\subsection{Survival Model}
\label{subsec:usermodel}
% There are several challenges in building a general and optimal learning-to-rank framework.
% One of the challenges is that it is infeasible to model the ranker of each document, since user's preference on documents is unstable and changing conditioned to different interaction. 
% Another challenge is position bias, caused by limitation of user propensity during browsing period, making item present at beginning more likely to be clicked.\par

% To address these challenges, it's natural to take both contextual information and position of each interaction into consideration and model preference with sequential browsing behavior modeling.\par

In the field of survival analysis \cite{li2016multi,ren2019deepa}, we investigate the probability of death event $z$ happening at each time.
Analogously, we here investigate the probability of click event $z$ happening at each document.
Let $z = i$ denote that event $z$ happens at $i$-th document $d_i$, and $z \geq i$ denote that event $z$ happens after $i$-th document $d_i$.
We then analyze the patient's investigation on underlying survival period, where a patient will keep \emph{`survival'} until she \emph{leave}s hospital or meets \emph{`death'}.
Actually, user behaviors on browsing are very similar, where a user will keep \emph{observ}ing until she \emph{leave}s due to lost of interest or \emph{click}s due to success in finding a worthwhile document.
Hence, we find that \emph{click} at each item corresponds to the \emph{`death'} status of one patient \cite{zhang2016bidaware}, and define \emph{click probability}, the probability density function (P.D.F.) of click occurring at $i$-th document $d_i$, as 
\begin{equation}
\begin{aligned}
p_i \doteq P(c_i = 1) \doteq  P(z = i) ,
\end{aligned}
\end{equation}
where $z$ denotes the position of clicked document.
Also, we see that \emph{observe} at each item corresponds to the \emph{`survival'} status of one patent \cite{zhang2016bidaware}.
Hence, we can derive the \emph{observe probability} at $i$-th document $d_i$ as the cumulative distribution function (C.D.F.), since user will keep browsing until she finds and clicks a favored one, as
\begin{equation}
\begin{aligned}
\label{eqn:discreteS}
S(i) \doteq P(o_i = 1) \doteq P( z \geq i) =  \sum_{\tau \geq i} P(z = \tau),
\end{aligned}
\end{equation}
which represents the probability of the click event occurring after document $d_i$,
% In other words, this formulation denotes the probability of browse log ending after $d_i$, 
\emph{i.e.}, probability of observing $d_i$.
Then it's straightforward to define the \emph{unobserve probability}, \emph{i.e.}, the probability of event occurring before the document $d_i$, as
\begin{equation}
\label{eqn:discreteW}
W(i) \doteq P(o_i = 0) \doteq P(z<i) = \sum_{\tau<i} P(z = \tau) .
\end{equation}
Hence, \emph{click probability} function at the $i$-th document can be calculated as
\begin{equation}
\label{eqn:discreteP}
\begin{aligned}
p_i & = P(z = i) = W(i+1) - W(i) \\
& = [1 - S(i+1)] - [1 - S(i)] \\
& = S(i) - S(i+1) .
\end{aligned}
\end{equation}
We define the \emph{relevance probability} as \emph{conditional click probability} according to Eq.~(\ref{eqn:relevance}), the click probability at document $d_i$ given that the previous document $d_{i-1}$ is observed, as
\begin{equation}
\label{eqn:discreteH}
h_i \doteq P(r_i = 1) =  \frac{P(c_i = 1)}{P(o_i = 1)} = \frac{P(z = i)}{P(z \geq i)} = \frac{p_i}{S(i)} ,
\end{equation}
which also means the probability that the click occurring document $z$ lies at $d_i$ given the condition that $z$ is larger than the last observation boundary.\par

For those non-click logs caused by user \emph{leave} behavior, we assume that user's favored document (\emph{i.e.}, \emph{click}) hides in the future session.
A similar scenario can be found in survival analysis when a patient \emph{leave}s hospital and finally meets \emph{`death'} sometime after investigation period.
Hence, we can regard these non-click logs as the censored clicked queries where censorship occurs in click.
Note that the data logs of unbiased learning-to-rank are represented as a set of triple $\{(\bm{x}, z, l)\}$, where $\bm{x}$ is the feature of the item and $l$ is the browse length.
Here $z$ is the position of clicked document $d_z$ if the user clicks in this browsing behavior, but $z$ is unknown (and we marked $z$ as null) in those non-click browsing histories.
Different from traditional causality models \cite{craswell2008experimental,chapelle2009dynamic}, survival model is able to capture observe patterns in both click and non-click queries.

\subsection{Deep Recurrent Survival Ranking Model}
\label{sec:rnn}
% Recall that we have presented the discrete user browsing model and discussed the click and unclick probability over the discrete user browsing space.
Based on survival model, we introduce our DRSR based on recurrent neural network $f_\theta$ with the parameter $\theta$, which captures the sequential patterns for \emph{conditional click probability} $h_i$ at every document $d_i$.
This structure also enables DRSR to take contextual information (\emph{i.e.}, observed documents) into consideration.
The detailed structure of DRSR is illustrated in Figure~\ref{fig:ranker}.
At each document $d_i$, the $i$-th RNN cell predicts the \emph{conditional click probability} $h_i$ given the document feature $\bm{x}_i$ as
\begin{equation}
\label{eqn:RNNH}
\begin{aligned}
h_i & = P(z = i \ | \ z \geq i, \bm{x}; \theta) = f_{\theta}(\bm{x}_i \ | \ b_{i-1}),
\end{aligned}
\end{equation}
where $f_\theta$ is the RNN function taking $\bm{x}$ as input and $h_i$ as output.
$b_{i-1}$ is the hidden vector calculated from the last RNN cell.
In our paper we implement the RNN function as a standard LSTM unit \cite{hochreiter1997long}, which has been widely used in sequence data modeling.\par

From Eqs.~(\ref{eqn:discreteW}), (\ref{eqn:discreteS}), (\ref{eqn:discreteH}) and (\ref{eqn:RNNH}), we can easily derive the \emph{observe probability} function $W(i)$ and the \emph{unobserve probability} function $S(i)$ as
\begin{equation}
\begin{aligned}
S(i | \bm{x}; \theta) & = P(i \leq z | \bm{x}; \theta) = P(z \neq 1, z \neq 2, \dots , z \neq i-1 | \bm{x}; \theta)\\
& = P(z \neq 1 | \bm{x}_1; \theta) \cdot P(z \neq 2 |z \neq 1, \bm{x}_2; \theta)  \cdots \\
& \cdot P(z \neq i-1|z \neq 1, \dots , z \neq i-2,\bm{x}_{i-1}; \theta)\\
& = \prod_{\tau:\tau < i} [1-P(z = \tau | z \geq \tau,\bm{x}_\tau; \theta)] = \prod_{\tau:\tau < i} (1-h_\tau) ,
\end{aligned}
\end{equation}
\begin{equation}
W(i|\bm{x};\theta) = P(i>z|\bm{x};\theta) = 1 - S(i|\bm{x}; \theta) = 1 - \prod_{\tau:\tau < i} (1-h_\tau) .
\end{equation}
% where $l_t$ is the time interval position for time $t$.
Here we use probability chain rule to calculate \emph{unobserve probability} $S(i)$ at the given document $d_i$ through multiplying the \emph{conditional unclick probability} $(1-h_\tau)$, \emph{i.e.}, inverse of the conditional click probability.\par

Moreover, taking Eqs.~(\ref{eqn:discreteP}) and (\ref{eqn:discreteH}) into consideration, the probability of clicked document $d_z$ directly lying at $d_i$, \emph{i.e.}, \emph{click probability} at document $d_i$ can written as
\begin{equation}
\label{eqn:RNNP}
p_i = P(z = i | \bm{x}; \theta) = h_{i} \prod_{\tau:\tau<i} (1 - h_\tau) .
\end{equation}

% Hence, we have introduced our DRL2R framework whose recurrent neural network in architecture and conditional probability in formulation can better capture user browsing behaviors through contextual information and sequential analysis.
% We then introduce point-wise loss function to mine click pattern in click queries and observe pattern in non-click queries under the framework of DRSR.

\begin{figure}[t]
	\vspace{-5mm}
	\centering
	\includegraphics[width=0.45\textwidth]{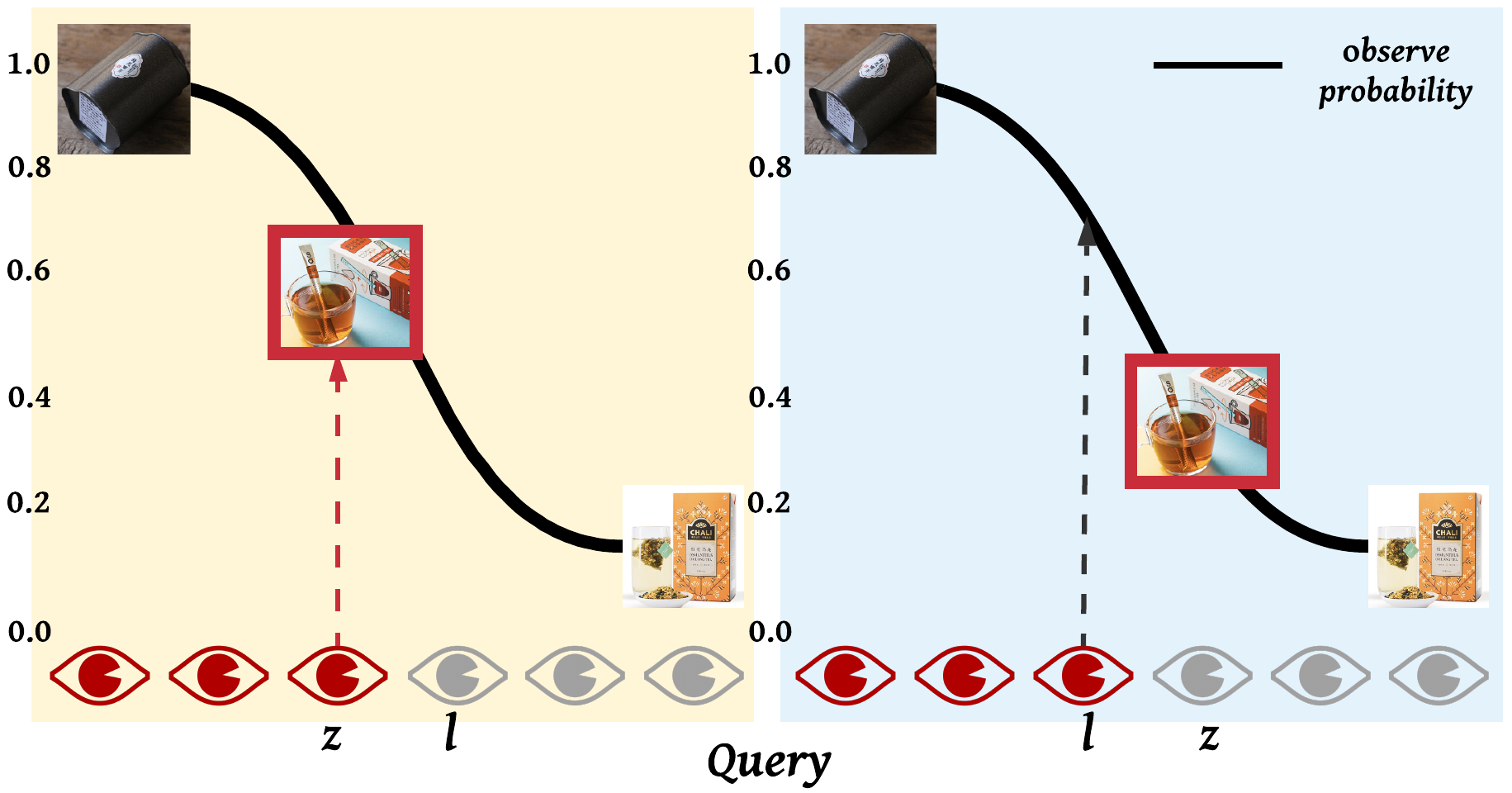}
	\vspace{-4mm}
	\caption {Intuition behind C.D.F. losses. The left and right sub-figures denote click and non-click cases respectively.}
	\label{fig:drsa}
	\vspace{-5mm}
\end{figure}

\subsection{Point-wise Loss Function}
\label{sec:pointwiseloss}
In point-wise setting, since there is no ground truth of either event probability distribution or relevance information, here we maximize the log-likelihood over the empirical data distribution to learn our deep model.\par
% Specifically, we introduce point-wise and pair-wise loss function in the following parts.

The first type of loss is based on the \emph{click probability} (P.D.F.) and it aims to minimize negative log-likelihood of the click document $d_j$ over the clicked logs as
\begin{equation}
\begin{aligned}
L_{\text{point}(z)} & = - \text{log} \prod_{(\bm{x}, z) \in \mathcal{D}_{\text{click}}} P(z = j | \bm{x}; \theta) =  - \text{log} \prod_{(\bm{x}, z) \in \mathcal{D}_{\text{click}}} p_j \\
& = - \text{log} \prod_{(\bm{x}, z) \in \mathcal{D}_{\text{click}}} [h_j \prod_{\tau: \tau<i} (1-h_\tau)] \\
& = - \sum_{(\bm{x}, z) \in \mathcal{D}_{\text{click}}} [\text{log} \ h_j + \sum_{\tau: \tau<i} \text{log}(1-h_\tau)] ,
\end{aligned}
\end{equation}
where $j$ is the position of true clicked document $d_j$ given the feature vector $\bm{x}$.\par

The second type of loss is based on the \emph{observe probability} (C.D.F.).
% Recall that there are click cases and non-click cases in the dataset. 
There are two motivations about the second loss corresponding to these two cases.
Let $z$ and $l$ represent clicked document position and browse length respectively. 
As is shown in Figure~\ref{fig:drsa}, the left sub-figure is the click case where $z$ has been known and $z \leq l$; The right sub-figure is the non-click case where $z$ is unknown (censored) but we only have the knowledge that $z > l$.\par

For the click cases as the left part of Figure~\ref{fig:drsa}, we need to ``push up'' the \emph{observe probability} for the document whose position is in range of $[0, l]$, while ``pull down'' the \emph{observe probability} for the document whose position is in range of $[l, \infty)$.
Thus, on one hand, we adopt the loss over the click cases that
\begin{equation}
\label{eqn:LossClick}
\begin{aligned}
L_{\text{click}} & = - \text{log} \prod_{(\bm{x}, l) \in \mathcal{D}_{\text{click}}} P(l \geq z|\bm{x};\theta) \\
& \approx  - \text{log} \prod_{(\bm{x}, l) \in \mathcal{D}_{\text{click}}} W(l|\bm{x};\theta) \\
& = - \sum_{(\bm{x},l) \in \mathcal{D}_{\text{click}}} \text{log} \ [1 - \prod_{\tau: \tau < l} (1 - h_\tau)] .
\end{aligned}
\end{equation}

As for the non-click cases in the right part of Figure~\ref{fig:drsa}, we just need to ``push up'' the \emph{observe probability} since we have no idea about true click document but we only know that $z > l$.
On the other hand, we just adopt the loss over the non-click dataset as
\begin{equation}
\label{eqn:LossUnclick}
\begin{aligned}
L_{\text{non-click}} & = - \text{log} \prod_{(\bm{x},l) \in \mathcal{D}_{\text{non-click}}} P(z > l| \bm{x}; \theta) \\
& \approx  - \text{log} \prod_{(\bm{x},l) \in \mathcal{D}_{\text{non-click}}} S(l|\bm{x};\theta) \\
& = - \sum_{(\bm{x},l) \in \mathcal{D}_{\text{non-click}}} \sum_{\tau: \tau < l} \text{log} \ (1-h_\tau) .
\end{aligned}
\end{equation}

\subsection{Permutation Document Model}
\label{sec:permutation}
\begin{figure}[h]
	\vspace{-3mm}
	\centering
	\includegraphics[width=0.3\textwidth]{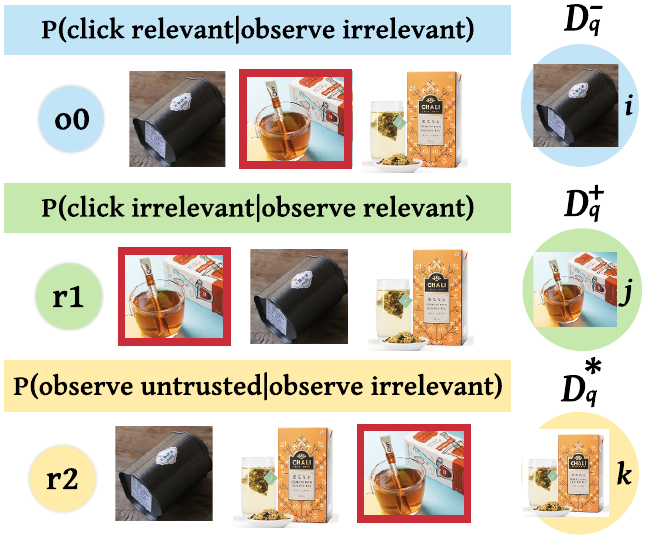}
	\vspace{-3mm}
	\caption {Illustration of permutation document modeling, where documents here are sampled from different subsets in the right side. Double sided arrow denotes exchange order operation.
	}
	\label{fig:orderless}
	\vspace{-5mm}
\end{figure}

Different from traditional pair-wise methods where binary classification accompanied with logistic regression is proposed to model relative relevance, as Figure~\ref{fig:orderless} shows, we here model the relative relevance via three conditional probabilities: \emph{click probability} (P.D.F.): (i) $P(z = j | z \geq i)$ for positive document $d_j$; (ii) $P(z = i | z \geq j)$ for negative document $d_i$; and \emph{observe probability} (C.D.F.): (iii) $P(z \geq k | z \geq i)$ for untrusted document $d_k$.
As Figure~\ref{fig:orderless} shows, the first one indicates the probability of user clicking document $d_j$ given she has browsed document $d_i$; 
the second one represents how likely user click document $d_i$ given she has observed document $d_j$; 
while the third one means the probability of user going on browsing document $d_k$ after observing document $d_i$.\par

It should be noted that only the first conditional probability is accessible to be measured since we can only obtain original order 0 (o1 in Figure~\ref{fig:orderless}).
In order to get rerank 1 (r1) and rerank 2 (r2), we need to permute the documents.
Recall that users often browse from the top to bottom, which may result in that higher the document ranked, more likely it to be clicked.
We are able to move clicked document $d_j$ forward since it will not change click behavior.
By this way, we obtain r1.
Considering there may exist relevant documents in those untrusted observations, we consider to move untrusted document $d_k$ forward to get r2.
Note that moving $d_k$ forward and $d_j$ backward may change click behavior, we here model observe probability instead.
This technique to mine more potential order based on an original order is inspired by XLNet \cite{yang2019xlnet}, so we call it permutation document modeling.\par

By this way, our model is able to 
(i) consider documents' display order in pair-wise setting by modeling relative relevance with conditional probability; 
(ii) take both trusted and untrusted observation into consideration;
(iii) conduct training procedure with more dependency pairs.

\subsection{Pair-wise Loss Functions}
\label{sec:pairwiseloss}
Different from point-wise loss functions, pair-wise loss functions can preserve relative information, \emph{e.g.}, relative relevance can be drawn from user's action to document $d_j$ given she has browsed document $d_i$.
There are three pair-wise loss functions for positive, negative and untrusted documents respectively based on the analysis in Section~\ref{sec:permutation}.\par
% Similar as point-wise loss above, we give pair-wise loss based on analysis in Section~\ref{sec:permutation} and framework in Section~\ref{sec:rnn}.
% These pair-wise functions are built for modeling positive, negative and untrusted documents respectively.
% Recall that we have proposed three factorization orders as illustrated in Figure~\ref{fig:orderless}.

In order to maximize log-likelihood of \emph{click probability} (P.D.F.): $P(z = j | z \geq i)$, \emph{i.e.}, clicking relevant document $d_j$ after observing irrelevant document $d_i$, we formulate the first pair-wise loss based on o0 query in Figure~\ref{fig:orderless} as
\begin{equation}
\begin{aligned}
L_{\text{pair}(o_0)} & = - \text{log} \prod_{(d_i, d_j) \in I_q} P(z = j| z \geq i,\bm{x};\theta) \\
& = - \text{log} \prod_{(d_i, d_j) \in I_q} \frac{P(z = j|\bm{x}; \theta)}{P(z \geq i)|\bm{x}; \theta)} = - \text{log} \prod_{(d_i, d_j) \in I_q} \frac{p_j}{S(i|\bm{x};\theta)} \\ 
& = - \sum_{(d_i, d_j) \in I_q} \{ [\text{log} \ h_j \sum_{\tau: \tau < j} \text{log} (1-h_\tau)] - \sum_{\tau: \tau<i}  \text{log} (1-h_\tau)\} .
\end{aligned}
\end{equation}
%% where $d_i$ and $d_j$ represent relevant and irrelevant documents corresponding to Figure~\ref{fig:pairwise} and Figure~\ref{fig:orderless}.
%% $L_{\text{pair}(o_0)}$ denotes the probability of user clicking on $d_i$ given she has browsed $d_j$, which describes the relevance difference based on conditional probability.
% Recall the assumption that document ranked top is more likely to be observed and clicked.
%We can rerank relevant document $d_j$ just after $d_i$, in order to avoid potential effects from other documents.\par
% In rerank 1 query, jointing considering relevant document $d_j$ ranking higher than irrelevant document $d_i$ and H2C assumption, it's unlikely for user to click on $d_i$.
In r1 query, we need to minimize log-likelihood of \emph{click probability} (P.D.F.): $P(z = i | z \geq j)$, \emph{i.e.}, clicking irrelevant document $d_i$ after observing relevant document $d_j$ and form the second pair-wise loss function as
\begin{equation}
\begin{aligned}
L_{\text{pair}(r_1)} & = \text{log} \prod_{(d_i, d_j) \in I_q} P(z = i| z \geq j,\bm{x};\theta) \\
& = \text{log} \prod_{(d_i, d_j) \in I_q} \frac{P(z = i|\bm{x}; \theta)}{P(z \geq j)|\bm{x}; \theta)} = \text{log} \prod_{(d_i, d_j) \in I_q} \frac{p_i}{S(j|\bm{x};\theta)} \\ 
& = \sum_{(d_i, d_j) \in I_q} \{ [\text{log} \ h_i \sum_{\tau: \tau < i} \text{log} (1-h_\tau)] - \sum_{\tau: \tau<j}  \text{log} (1-h_\tau)\} .
\end{aligned}
\end{equation}
% 	\kan{negative loss?}

% Although the untrusted observations, a.k.a. uncertainly negative documents do not contain definite information about the click event time, we would only know that relevance of these documents is often considered between relevant and irrelevant ones. 
%% These two loss functions focus on modeling relevance difference between irrelevant document $d_i$ and relevant document $d_j$ and ignore to take uncertainly irrelevant document $d_k$ into consideration.
%% This is mainly because we don't capture enough information for $d_k$.

For r2 query, we measure the relative relevance between trusted (\emph{i.e.}, relevant and irrelevant) and untrusted documents via user observe behavior.
Specially, we evaluate \emph{observe probability} (C.D.F.): $P(z \geq k | z \geq i)$, \emph{i.e.}, probability of user going browsing $d_k$ after she has observed $d_i$ as
\begin{equation}
\begin{aligned}
L_{\text{pair}(r_2)} & = - \text{log} \prod_{(d_i,d_k) \in I_q} P(z \geq k| z \geq i, \bm{x}; \theta) \\
& = - \text{log} \prod_{(d_i,d_k) \in I_q} \frac{P(z \geq k|\bm{x};\theta)}{P(z \geq i|\bm{x}; \theta)} = - \text{log} \prod_{(d_i,d_k) \in I_q} \frac{S(k|\bm{x};\theta)}{S(i|\bm{x};\theta)} \\
& = - \sum_{(d_i, d_k) \in I_q}[\sum_{\tau:\tau < k} \text{log} \ (1-h_\tau) - \sum_{\tau:\tau < i} \text{log} \ (1-h_\tau)] .
\end{aligned}
\end{equation}
% Also we can rerank uncertainly irrelevant $d_k$ just after $d_i$ to avoid potential effects from other documents.\par

\vspace{-3mm}
\subsection{Model Realization}
In this section, we unscramble some intrinsic properties of our deep model and analyze the model efficiency in this section.\par

\minisection{Properties of Loss Function} First of all, we take the view of click prediction of our methodology.
As is known that there is a click status, \emph{i.e.}, an indicator of click event, for each sample as
\begin{equation}
\omega = \left\{
\begin{aligned}
1 \ & \ \text{if} \ l \geq z\\
0 \ & \ \text{otherwise} \  l < z .
\end{aligned}
\right.
\end{equation}

% For the clicked logs, each sample ($\bm{x}$, $z$, $l$) is uncensored (i.e., $z$ is known) where $\omega = 1$; while for the non-clicked data, the true clicked document $d_z$ is unknown but the we only have the idea that $z \geq l$, so that $\omega = 0$.\par

Hence, taking Eqs.~(\ref{eqn:LossClick}) and (\ref{eqn:LossUnclick}) altogether and we may find that combination of $L_{\text{click}}$ and $L_{\text{non-click}}$ describes the classification of click status at document $d_l$ of each sample as
\begin{equation}
\begin{aligned}
L_2 & = L_{\text{click}} + L_{\text{non-click}} \\
& = - \text{log} \prod_{(\bm{x},l) \in \mathcal{D}_{\text{click}}} P(l \geq z|\bm{x};\theta) - \text{log} \prod_{(\bm{x},l) \in \mathcal{D}_{\text{non-click}}} P(z > l|\bm{x}; \theta) \\
& \approx - \text{log} \prod_{(\bm{x},l) \in \mathcal{D}} [W(l|\bm{x};\theta)]^\omega \cdot [1-W(l|\bm{x};\theta)]^{1-\omega} \\
& = - \sum_{(\bm{x},l) \in \mathcal{D}} \{ \omega \cdot \text{log} \ W(l|\bm{x};\theta) + (1-\omega) \cdot \text{log} \ [1 - W(l|\bm{x};\theta)] \} ,
\end{aligned}
\end{equation}
which is the cross entropy loss for predicting click status at time $t$ given $\bm{x}$ over all the data $\mathcal{D} = \mathcal{D}_{\text{click}} \cup \mathcal{D}_{\text{non-click}}$ .\par

Combining all the objective functions and our goal is to minimize the negative log-likelihood over all the data samples including both clicked and non-click data as
\begin{equation}
\label{eqn:LossAll}
\begin{split}
& \text{arg} \mathop{\text{min}}_\theta \alpha L_1 + (1-\alpha)L_2 \\
& \text{where} \ L_1 = \left\{
\begin{aligned}
& L_{\text{point}(z)} & \text{point-wise} \\
& L_{\text{pair}(o_0)} + L_{\text{pair}(r_1)} + L_{\text{pair}(r_2)} & \text{pair-wise} ,
\end{aligned}
\right.
\end{split}
\end{equation}
where the hyper-parameter $\alpha$ controls the order of magnitudes of the gradients from the two losses at the same level to stabilize the model training.\par

In the traditional and related works, they usually adopt only $L_1$ based on \emph{click probability} (P.D.F.) in unbiased learning-to-rank field \cite{hu2019unbiased} for click prediction or $L_{\text{non-click}}$ in survival analysis field \cite{cox1972regression,katzman2018deepsurv} for censorship handling.
We propose a comprehensive loss function which learns from both click logs and non-click logs.
From the discussion above, $L_{\text{click}}$ and $L_{\text{non-click}}$ collaboratively learns the data distribution from the \emph{observe probability} (C.D.F.) view.

\minisection{Model Efficiency} 
% Here we analyze the computational complexity of deep unbiased learning-to-rank.
As shown in Eq.~(\ref{eqn:RNNH}), each recurrent unit $f_\theta$ takes ($\bm{x}$, $z_l$, $b_{l-1}$) as input and outputs probability scalar $h_l$ and hidden vector $b_l$ to the next unit.
Let $L$ be the maximal browse length, so the calculation of the recurrent units will run for maximal $L$ times.
We assume the average case time performance of recurrent units $f_\theta$ is $O(C)$, which is related to the implementation of the unit \cite{zhang2016architectural}, recurrent skip coefficients, yet can be paralleled through GPU processor.
The subsequent calculation is to obtain the multiplication results of $h_l$ or $(1-h_l)$ to get the results of $p_z$ and $S$, as that in Figure~\ref{fig:ranker}, whose complexity is $O(L)$.
Thus the overall time complexity is $O(CL)+O(L)=O(CL)$, which is the same as the original recurrent neural network model.\par

\vspace{-2mm}
\subsection{Learning Algorithm}
We provide the learning algorithm of DRSR in Algorithm~\ref{algo:framework}.
It should be noted that we calculate loss function $L(\theta)$ with $P(c=1)$ in training procedure since only click data is available; while we use $P(r=1)$ in inference procedure, which is actually the core of unbiased learning-to-rank. 
% The input is a dataset $\mathcal{D} = {(\bm{x}, z, l)}$, containing click dataset $\mathcal{D}_{\text{click}}$ and non-click dataset $\mathcal{D}_{\text{non-click}}$, together with hyper-parameter.
% The output is an unbiased ranker $f_\theta$ with parameter $\theta$.
% As is outlined in Algorithm~\ref{algo:framework}, DRL2R iteratively calculates \emph{relevance probability} $P(o = 1 | \bm{x})$ and \emph{click probability} $P(c = 1 | \bm{x})$ in line 5, computes the corresponding loss $L(\theta)$ in either point-wise or pair-wise setting in line 7 and trains the model in line 8.

\vspace{-2mm}
\begin{algorithm}[h!]
	\caption{Deep Recurrent Survival Ranking (DRSR)}
	\label{algo:framework}
	\begin{algorithmic}[1]
		\REQUIRE
		dataset $\mathcal{D} = \mathcal{D}_{\text{click}} \cup \mathcal{D}_{\text{non-click}}$;
		% hyper-parameter $\alpha$
		\ENSURE
		unbiased ranker $f_\theta$ with parameter $\theta$
		\vspace{1mm}
		\STATE Initialize all parameters.
		\REPEAT
		\STATE Randomly sample a batch $\mathcal{B}$ from $\mathcal{D}$
		\FOR {each point $d_q$ or pair $(d_i, d_j) \in I_q$ in $\mathcal{B}$}
		\STATE Calculate $P(r = 1)$ and $P(c = 1)$ using Eqs.~(\ref{eqn:RNNH}) and (\ref{eqn:RNNP}).
		%	\STATE Re-rank according to $p_q$	
		\ENDFOR
		\STATE Compute corresponding $L(\theta)$ according to Eq.~(\ref{eqn:LossAll}).
		\STATE Update $\theta$ by minimizing $L(\theta)$.
		\UNTIL convergence
	\end{algorithmic}
\end{algorithm}
\vspace{-5mm}

\section{Experiments}
\label{sec:exp}
In this section, we present the experimental setup and the corresponding results under various evaluation metrics.
% with significance test.
Furthermore, we look deeper into our model and analyze some insights of the experiment results.
Moreover, we have also published our code$^1$.\blfootnote{$^1$Reproducible code link: \url{https://github.com/Jinjiarui/DRSR}.}
We start with three reasearch questions (RQ) to lead the experiments and the following discussions.

\begin{itemize}[topsep = 3pt,leftmargin =5pt]
	\item (\textbf{RQ1}) Compared with the baseline models, does DRSR achieve state-of-the-art performance in unbiased learning-to-rank?
	\item (\textbf{RQ2}) Are debiased method, \emph{i.e.}, survival model, truely necessary for improving performance? 
	\item (\textbf{RQ3}) Can DRSR learn robustly under different situtation, \emph{i.e.}, simulation, bias degree and number of data?
\end{itemize}

\subsection{Datasets and Experiment Flow}
We conduct our experiment on Yahoo search engine dataset named Yahoo! learning-to-rank challenge dataset and Alibaba recommender system dataset.
We choose NDCG at position 1, 3, 5 and MAP as evaluation measures in relevance ranking.
% which has been published in \cite{chapelle2011yahoo}, contains 29921 queries and 710k documents.
% Each query document pair is represented by
% There is no click data associated with the Yahoo dataset.
% We follow the procedure in \cite{ai2018unbiasedb} to generate synthetically click data from Yahoo dataset$^\dagger$ for offline evaluation.

\begin{table*}[t]
	\centering
	\caption{Comparison of different unbiased learning-to-rank methods under Yahoo Search Engine and Alibaba Recommender System.
		CCM is utilized as click generation model.
		% Significant improvements or degradations with respect to DRSR with Pointwise Debiasing are indicated with +/-.
		% We represent p-value with the number of +/-, \emph{i.e.}, 
		* indicates p-value < 0.001 in significance test vs the best baseline.}
	\vspace{-3mm}
	\resizebox{0.95\textwidth}{!}{
		\begin{tabular}{|c|c|c|c|c|c|c|c|c|c|}
			\hline
			\multirow{2}{*}{Ranker} & \multirow{2}{*}{Debiasing Method} & \multicolumn{4}{c|}{Yahoo Search Engine (CCM)} & \multicolumn{4}{c|}{Alibaba Recommender System (CCM)} \\
			\cline{3-10}
			{} & {} & MAP & NDCG@1 & NDCG@3 & NDCG@5 & MAP & NDCG@1 & NDCG@3 & NDCG@5 \\
			\hline
			\multirow{5}{*}{DRSR (Ours)} & Labeled Data & 0.861 & 0.747 & 0.759 & 0.771 & 0.850 & 0.737 & 0.741 & 0.755 \\
			\cline{2-10}
			{} & Pairwise Debiasing & \textbf{0.842}$^{*}$ & \textbf{0.719}$^{*}$ & \textbf{0.721}$^{*}$ & \textbf{0.737}$^{*}$ & \textbf{0.831}$^{*}$ & \textbf{0.684}$^{*}$ & \textbf{0.685}$^{*}$ & \textbf{0.707}$^{*}$\\
			\cline{2-10}
			{} & Pointwise Debiasing & \textbf{0.839}$^{*}$ & \textbf{0.713}$^{*}$ & \textbf{0.717}$^{*}$ & \textbf{0.730}$^{*}$ & \textbf{0.830}$^{*}$ & \textbf{0.682}$^{*}$ & \textbf{0.684}$^{*}$ & \textbf{0.706}$^{*}$\\
			\cline{2-10}
			{} & Regression-EM \cite{wang2018position} & 0.829 & 0.679 & 0.685 & 0.701 & 0.820 & 0.657 & 0.668 & 0.673\\
			\cline{2-10}
			{} & Click Data & 0.817 & 0.636 & 0.652 & 0.667 & 0.810 & 0.613 & 0.627 & 0.658 \\
			\hline
			\multirow{4}{*}{LambdaMART} & Labeled Data & 0.854 & 0.745 & 0.745 & 0.757 & 0.847 & 0.729 & 0.732 & 0.743\\
			\cline{2-10}
			{} & Ratio Debiasing \cite{hu2019unbiased} & 0.830 & 0.688 & 0.685 & 0.699 & 0.821 & 0.661 & 0.669 & 0.674\\
			\cline{2-10}
			{} & Regression-EM \cite{wang2018position} & 0.826 & 0.669 & 0.676 & 0.691 & 0.818 & 0.636 & 0.651 & 0.667 \\
			\cline{2-10}
			{} & Click Data & 0.813 & 0.628 & 0.646 & 0.673 & 0.804 & 0.603 & 0.618 & 0.646 \\
			\hline
			\multirow{4}{*}{DNN} & Labeled Data & 0.831 & 0.677 & 0.685 & 0.705 & 0.824 & 0.674 & 0.679 & 0.693 \\
			\cline{2-10}
			{} & Dual Learning Algorithm \cite{ai2018unbiaseda} & 0.825 & 0.672 & 0.678 & 0.691 & 0.814 & 0.629 & 0.647 & 0.674 \\
			\cline{2-10}
			{} & Regression-EM \cite{wang2018position} & 0.823 & 0.665 & 0.669 & 0.687 & 0.813 & 0.628 & 0.645 & 0.672\\
			\cline{2-10}
			{} & Click Data & 0.809 & 0.611 & 0.619 & 0.648 & 0.801 & 0.600 & 0.612 & 0.641\\
			\hline
		\end{tabular}
	}
	\label{tb:res}
	\vspace{-3mm}
\end{table*}

\minisection{Data Description}
Two large-scale real-world datasets$^2$ \blfootnote{$^2$Dataset download link: \url{http://webscope.sandbox.yahoo.com}.} are used in our experiments.
\begin{itemize}[topsep = 3pt,leftmargin =5pt]
	\item \textbf{Yahoo search engine dataset} is one of the largest benchmark dataset widely used in unbiased learning-to-rank \cite{hu2019unbiased,ai2018unbiaseda}.
	It consists of 29,921 queries and 710k documents. 
	Each query document pair is represented by a 700-dimensional feature vector manually assigned with a label denoting relevance at 5 levels \cite{chapelle2011yahoo}.
	
	\item \textbf{Alibaba recommender system dataset} is a proprietary dataset  where we regard each user and item information as query and document feature respectively.
	Similar approaches can be found in  \cite{wang2006user,zhu2018learning}.
	It contains 178,839 queries and 3,133k documents.
	Each query document pair is represented by a 1750-dimensional feature vector and assigned with a binary label denoting relevance.	
\end{itemize}

\minisection{Click Data Generation}
The click data generation process in \cite{ai2018unbiaseda,hu2019unbiased} is followed.
First, one trains a Rank SVM model using $1\%$ of the training data with relevance labels.
Next, one uses trained model to create an initial ranking list for each query. 
Then, one simulates user browsing process and samples clicks from the initial list.
We utilize two simulation model here:

\begin{itemize}[topsep = 3pt,leftmargin =5pt]
	\item \textbf{PBM} \cite{richardson2007predicting} simulates this user browsing behavior based on the assumption that the bias of a document only depends on its position, which can be formulated as $P(o_i) = \rho_i^\tau$, where $\rho_i$ represents position bias at position $i$ and $\tau \in [0, +\infty]$ is a parameter controlling the degree of position bias.
	The position bias $\rho_i$ is obtained from an eye-tracking experiment in \cite{joachims2005accurately} and the parameter $\tau$ is set as 1 by default. 
	It also assumes that a user decides to click a document $d_i$ according to probability $P(c_i) = P(o_i) \cdot P(r_i)$.
	
	\item \textbf{CCM} \cite{guo2009click} is a cascade model, which assumes that the user browses the search results in a sequential order from top to bottom.
	The user browse behaviors are both conditioned on current and past documents, as
	$P(c_i = 1 | o_i = 0) = 0$,
	$P(c_i = 1 | o_i = 1, r_i) = P(r_i)$,
	$P(o_{i+1} = 1 | o_i = 0) = 0$,
	$P(o_{i+1} = 1 | o_i = 1, c_i = 0) = \upgamma_1$,
	$P(o_{i+1} = 1 | o_i = 1, c_i = 1, r_i) = \upgamma_2 \cdot (1-P(r_i)) + \upgamma_3 \cdot P(r_i)$.
	The parameter is obtained from experiment in \cite{guo2009click} with $\upgamma_2=0.10$ and $\upgamma_3=0.04$ for navigational queries (Yahoo search engin); $\upgamma_2=0.40$ and $\upgamma_3=0.27$ for informational queries (Alibaba recommender system).
	$\upgamma_1$ is set as 0.5 by default.
\end{itemize}

The probability of relevance $P(r_i)$ is calculated by $P(r_i) = \epsilon + (1-\epsilon) \cdot \frac{2^{y_i} - 1}{2^{y_{\text{max}}} - 1}$, where $y_i \in [0, 4]$ represents relevance level.
The parameter $\epsilon$ denotes click noise and is set as 0.1 as default.\par

\vspace{-3mm}
\subsection{Compared Settings}
We made comprehensive comparisons between our model and the baselines.
The baselines are created by combining the learning-to-rank algorithm with the state-of-the-art debiasing methods.
\begin{itemize}[topsep = 3pt,leftmargin =5pt]
	\item \textbf{Regression-EM}: \citet{wang2018position} proposed regression-based EM method where position bias is estimated directly from regular production clicks.
	% The algorithm incorporates with various ranking models through implementing those ranker in M step.
	\item \textbf{Dual Learning Algorithm}: \citet{ai2018unbiaseda} proposed a dual learning which can jointly learn a ranker and conduct debiasing of click data.
	% The algorithm implements both the ranking model and the debiasing model with unbiased propensity estimation.
	\item \textbf{Ratio Debiasing}: \citet{hu2019unbiased} proposed an unbiased pair-wise learning-to-rank based on inverse propensity weight (IPW) \cite{wang2016learning}.
	\item \textbf{Point-wise Debiasing}: Our proposed debiasing method DRSR which is adapted in point-wise setting.
	\item \textbf{Pair-wise Debiasing}: Our proposed debiasing method DRSR which is adapted in pair-wise setting.
	\item \textbf{Click Data}: We utilize the raw click data without debiasing to train the ranker, whose performance is regarded as a lower bound.
	\item \textbf{Labeled Data}: The human annotated relevance labels without any bias are used as data for training the ranker and we consider its performance as an upper bound. 
\end{itemize}

There are several learning-to-rank algorithms cooperated with debiasing methods.
\begin{itemize}[topsep = 3pt,leftmargin =5pt]
	\item \textbf{DRSR}: Our model.
	\item \textbf{DNN}: A deep neural network as described in \cite{ai2018unbiaseda} is implemented as a ranker.
	% We adopt the code provided by Ai et al.$^\S$.\blfootnote{$^\S$\url{https://github.com/QingyaoAi/Dual-Learning-Algorithm-for-Unbiased-Learning-to-Rank}}  
	\item \textbf{LambdaMART}: We implement Unbiased LambdaMART by modifying the LambdaMART tool in LightGBM \cite{ke2017lightgbm}.
	% We utilize the code provided by Hu et al.$^{\i}$\blfootnote{$^{\i}$\url{https://github.com/acbull/Unbiased_LambdaMart}}
\end{itemize}

In summary, there are 11 baselines to compare with our model.
Note that Dual Learning and DNN are tightly coupled.
Also, the same situation happens in Ratio Debiasing and LambdaMART. 
We don't combine Ratio Debiasing with DNN and Dual Learning with LambdaMART, as it's beyond the scope of this paper.

\begin{figure*}[t]
	\centering
	\includegraphics[width=\textwidth]{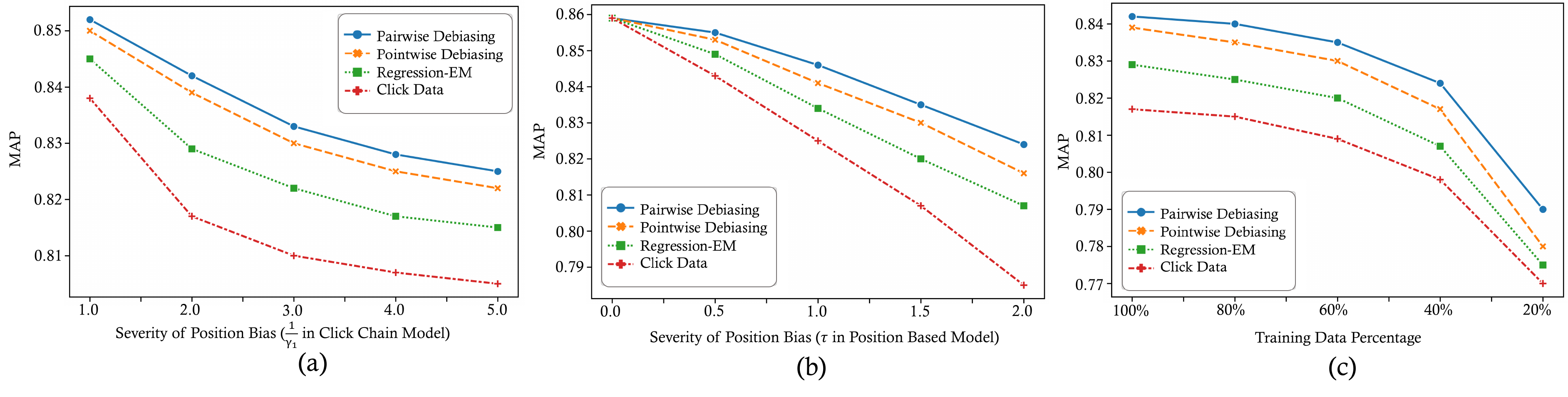}
	\vspace{-7mm}
	\caption {(a) \& (b) Performance of DRSR against other debiasing methods with different degrees of position bias. (c) Performance of DRSR against other debiasing methods with different sizes of training data.}
	\vspace{-3mm}
	\label{fig:data}
\end{figure*}

\vspace{-3mm}
\subsection{Result Analysis}
\label{sec:res}

\vspace{-2mm}
\minisection{Experimental Results and Analysis (RQ1)}
Table~\ref{tb:res} summerizes the results.
We see that our method of Deep Recurrent Survival Ranking Models (DRSR + Pointwise Debiasing / Pairwise Debiasing) significantly outperform all the other baseline methods.
The results of Ratio Debiasing, Regression-EM and Dual Learning Algorithm are compariable with those reported in the original papers.
In particular, we have the following findings:
\begin{itemize}[topsep = 3pt,leftmargin =5pt]
	\item  Our models based DRSR achieve better performances than all the state-of-the-art methods in terms of all measures, which indicates our framework DRSR outperforms other models such as LambdaMART and DNN.
	The reason seems to be that our framework enables to find correlation of user various behaviors and mine hidden observe pattern in non-click queries.
	\item Pairwise Debiasing works better than the other debiasing methods when combined with DRSR framework.
	This implies that considering relative relations between trusted and untrusted observation can enhance model performance.
	\item When conducted in Alibaba Recommender System, the performances of all models decrease significantly.
	This implies that items in recommendation are in a larger scale and have a more complex feature space than in search engine.
	\item The performances of Pairwise Debiasing and Pointwise Debiasing get closer in Alibaba Recommender System.
	This indicates that it is challenging to define and capture relative relevance in recommendation, since various items in different categories can be displayed at the same time.
	Also, the user preference is more personalized, dynamic and even noisy.
\end{itemize}

\begin{figure}[b]
	\centering
	\vspace{-5mm}
	\includegraphics[width=0.35\textwidth]{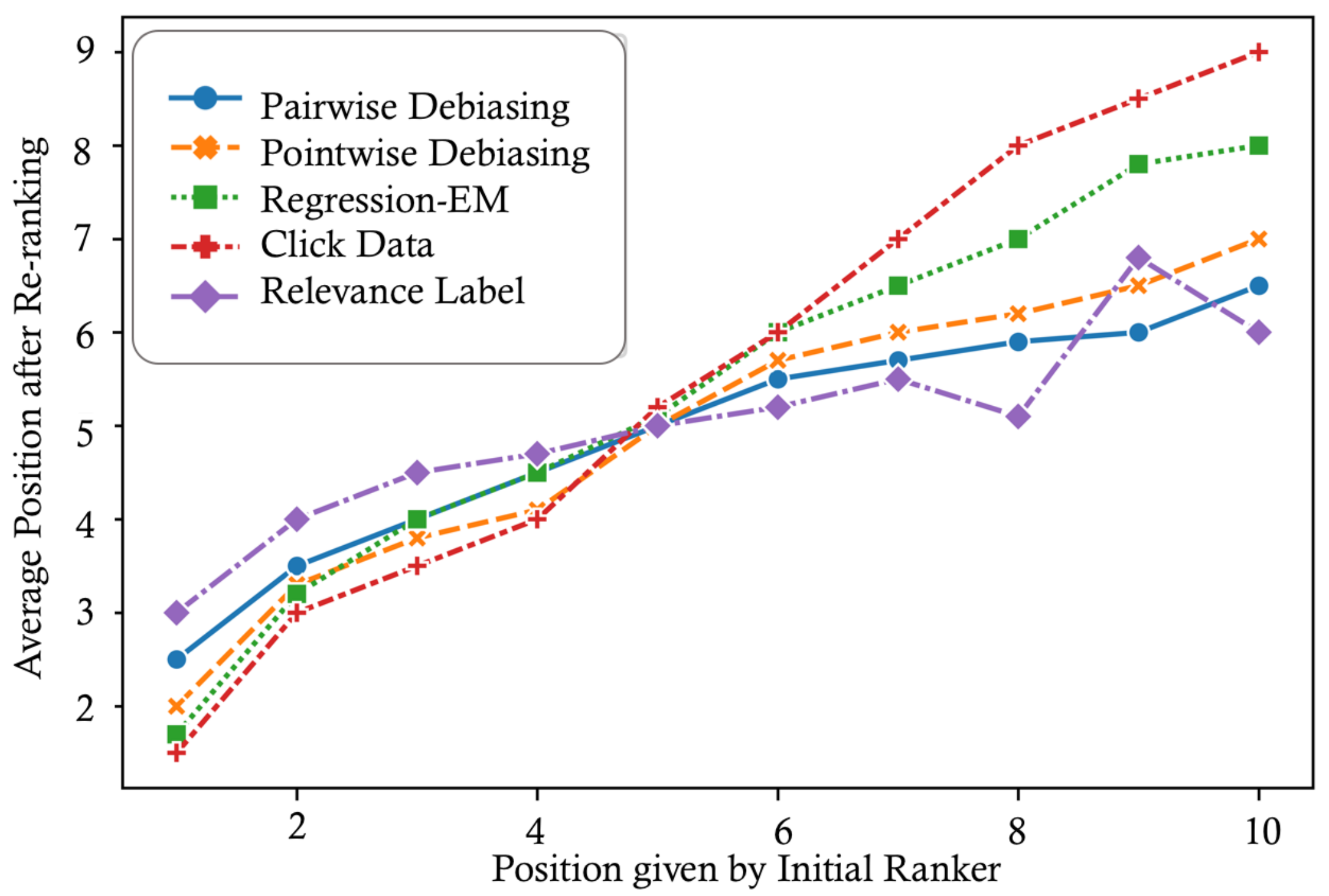}
	\vspace{-3mm}
	\caption {Average positions after re-ranking of documents at each original position by different debiasing methods combined with DRSR.}
	\label{fig:position}
	\vspace{-3mm}
\end{figure}

\minisection{Ablation Study (RQ2)}
In order to analyze the importance of survival model (debiasing method) and recurrent neural network (ranker), we also conduct experiment of recurrent neural network without survival model and summarize results as Click Data of DRSR in Table~\ref{tb:res}.
We can find that sophisticated algorithms like DRSR and LambdaMART are sensitve to position bias when comparing the performance of Click Data with Debiasing methods, which indicates the significance for unbiased learning-to-rank.
Also, we can see that when trained with human labeled data, DRSR achieves the best performance (Labeled Data) which implies that there is still much room for improvement in unbiased learning-to-rank.

\minisection{Visualization Analysis (RQ2)}
We investigated whether the performance improvement by DRSR is indeed from reduction of position bias through comparing the ranking list given by the initial ranker with debiased ranker.\par

We first identified the documents at each position given by the initial ranker.
Then, we calculated the average positions of the documents at each original position after re-ranking by various debiasing methods, combined with DRSR.
We also calculated the average positions of the documents after re-ranking by their relevance labels, which is regarded as the ground truth.
Ideally, the average positions by the debiasing methods should get close to the average position by the relevance labels.
We summarized the results and show them in Figure~\ref{fig:position}.\par

One can see that the curve of Click Data (in red) is away from that of Relevance Label (in purple), indicating that directly using click data without debiasing can be problematic.
The curve of Pairwise Debiasing (in blue) and Pointwise Debiasing (in orange) are the closest  to the curve of Relevance Label, representing that the performance enhancement of DRSR is indeed from effective debiasing.

\begin{table}[t]
	\centering
	\caption{Comparison with PBM as click generation model. Notations are same with Table~\ref{tb:res}.}
	\vspace{-3mm}
	\resizebox{0.45\textwidth}{!}{
		\begin{tabular}{|c|c|c|c|c|}
			\hline
			\multicolumn{5}{|c|}{Yahoo Search Engine (PBM)} \\
			\hline
			Ranker & MAP & NDCG@1 & NDCG@3 & NDCG@5 \\
			\hline
			\multirow{5}{*}{DRSR (Ours)} & 0.861 & 0.747 & 0.759 & 0.771 \\
			\cline{2-5}
			{} & \textbf{0.848}$^{*}$ & \textbf{0.726}$^{*}$ & \textbf{0.737}$^{*}$ & \textbf{0.745}$^{*}$\\
			\cline{2-5}
			{} & \textbf{0.843}$^{*}$ & \textbf{0.723}$^{*}$ & \textbf{0.731}$^{*}$ & \textbf{0.740}$^{*}$\\
			\cline{2-5}
			{} & 0.834 & 0.698 & 0.705 & 0.712\\
			\cline{2-5}
			{} & 0.825 & 0.671 & 0.679 & 0.693 \\
			\hline
			\multirow{4}{*}{LambdaMART} & 0.854 & 0.745 & 0.745 & 0.757\\
			\cline{2-5}
			{} & 0.836 & 0.717 & 0.716 & 0.728 \\
			\cline{2-5}
			{} & 0.830 & 0.685 & 0.684 & 0.700 \\
			\cline{2-5}
			{} & 0.820 & 0.658 & 0.669 & 0.672 \\
			\hline
			\multirow{4}{*}{DNN} & 0.831 & 0.677 & 0.685 & 0.705 \\
			\cline{2-5}
			{} & 0.828 & 0.674 & 0.683 & 0.697 \\
			\cline{2-5}
			{} & 0.829 & 0.676 & 0.684 & 0.699\\
			\cline{2-5}
			{} & 0.819 & 0.637 & 0.651 & 0.667\\
			\hline
		\end{tabular}
	}
	\vspace{-7mm}
	\label{tb:rec}
\end{table}

\minisection{Generalizability Analysis (RQ3)}
The click data utilized in Table~\ref{tb:res} is generated by Click Chain Model (CCM), a cascade model, which assumes that the user browses the search results in a sequential order.
One can see that setting of DRSR and CCM match each other well, which may affect the performance.
Hence, we need to evaluate DRSR in a more general view.
The Position Based Model (PBM) assumes that the bias of a document only depends on its position, which is a general approximation of user click behavior in practice.
We compared the same baseline methods here.
Again, we found that DRSR significantly outforms the baselines, indicating that our model is indeed an effective method.

\minisection{Robustness Analysis (RQ3)}
We further evaluated the robustness of DRSR under different degrees of position bias and different size of training data.
In the above experiments, we only tested the performance of DRSR with click data generated from a single click model, \emph{i.e.}, $\upgamma_1=0.5$ in Click Chain Model and $\tau=1$ in Position Based Model.
Here, $\upgamma_1$ and $\tau$ influence the probability that user exams the next result.
Obviously, the smaller $\upgamma_1$ and larger $\tau$ indicate that the user will have a smaller probability to continue reading, which means a more severe position bias.
Therefore, here we set the two hyper-parameters
to different values and examined whether DRSR can still work equally well.\par

Figure~\ref{fig:data}(a) \& (b) show the results in terms of MAP with different degrees of position bias.
The results in terms of other measures have similar trends. 
When $\tau$ in PBM equals 0, there is no position bias; while $\upgamma_1$ in CCM equals 1, there still exist position bias brought from $\upgamma_2$ and $\upgamma_3$.
The results of all debiasing methods are similar to that of using click data only.
As we add more position bias, \emph{i.e.}, $\tau$ increases and $\upgamma_1$ decreases, the performances of all the debiasing methods decrease dramatically.
However, under all settings DRSR can get less affected by position bias and consistently maintain the best results.
This indicates that DRSR is robust to different
degrees of position bias.
\par

Next, we investigated the robustness of DRSR under different sizes of training data.
We first randomly selected a subset of training data (\emph{i.e.}, 20\% - 100\%) to generate different sizes of click datasets, and then used these datasets to evaluate the performances of DRSR with different debiasing methods.
To make fair comparsion, we utilized the same subsets of training data for Regression-EM.\par

As shown in Figure~\ref{fig:data}(c), when the size of training data decreases, the improvements obtained by the debiasing methods also decrease.
The reason seems to be that the position bias estimated from insufficient training data is not accurate, which can hurt the performances of the debiasing methods. 
DRSR Debiasing which adopts a joint training mechanism, can still achieve the best performances in such cases. 
Also, Pairwise Debiasing enhances its performance via data augmentation in document permutation model.
% This result is in accordance with the observation reported in \cite{joachims2017unbiased} that simply increasing the amount of biased training data cannot help build a reliable ranking model, but after debiasing it is possible to learn a better ranker with more training data. 
The experiment shows that DRSR can still work well even with limited training data.

\section{Conclusion and Future Work}
\label{sec:conclusion}
In this paper, we propose an innovative framework named DRSR where we adopt survival analysis techniques accompanied with probability chain rule to derive the joint probability of user various behaviors.
This framework enables unbiased model to leverage the contextual information in the ranking list to enhance the performance.
Also, we incorporate with survival analysis, and thus can model the non-click queries as the censored click logs, where the censorship occurs in click.
We design a novel objective function to mine the rich observe and click patterns hidden in both click and non-click queries.
Also, we extend pair-wise loss to capture relative relevance between trusted feedback and untrusted observation via conditional probability.
In the future work, it would be interesting to investigate better solution to model multiple-click session and consider good and bad in abandoned, \emph{i.e.}, non-click queries, respectively.

\minisection{Acknowledgments} 
The co-corresponding authors are Weinan Zhang and Kan Ren.
We thank the support of National Natural Science Foundation of China (Grant No. 61702327, 61772333, 61632017) and Wu Wen Jun Honorary Doctoral Scholarship from AI Institute, Shanghai Jiao Tong University.

% \clearpage
\bibliographystyle{ACM-Reference-Format}
\balance
\bibliography{pos-bias}


\begin{thebibliography}{47}

%%% the Web address.

\ifx \showCODEN    \undefined \def \showCODEN     #1{\unskip}     \fi
\ifx \showDOI      \undefined \def \showDOI       #1{#1}\fi
\ifx \showISBNx    \undefined \def \showISBNx     #1{\unskip}     \fi
\ifx \showISBNxiii \undefined \def \showISBNxiii  #1{\unskip}     \fi
\ifx \showISSN     \undefined \def \showISSN      #1{\unskip}     \fi
\ifx \showLCCN     \undefined \def \showLCCN      #1{\unskip}     \fi
\ifx \shownote     \undefined \def \shownote      #1{#1}          \fi
\ifx \showarticletitle \undefined \def \showarticletitle #1{#1}   \fi
\ifx \showURL      \undefined \def \showURL       {\relax}        \fi
% invisible to TeX
\providecommand\bibfield[2]{#2}
\providecommand\bibinfo[2]{#2}
\providecommand\natexlab[1]{#1}
\providecommand\showeprint[2][]{arXiv:#2}

\bibitem[\protect\citeauthoryear{Agarwal, Takatsu, Zaitsev, and
  Joachims}{Agarwal et~al\mbox{.}}{2019}]%
        {agarwal2019general}
\bibfield{author}{\bibinfo{person}{Aman Agarwal}, \bibinfo{person}{Kenta
  Takatsu}, \bibinfo{person}{Ivan Zaitsev}, {and} \bibinfo{person}{Thorsten
  Joachims}.} \bibinfo{year}{2019}\natexlab{}.
\newblock \showarticletitle{A General Framework for Counterfactual
  Learning-to-Rank}. In \bibinfo{booktitle}{\emph{SIGIR}}.
\newblock


\bibitem[\protect\citeauthoryear{Ai, Bi, Luo, Guo, and Croft}{Ai
  et~al\mbox{.}}{2018a}]%
        {ai2018unbiaseda}
\bibfield{author}{\bibinfo{person}{Qingyao Ai}, \bibinfo{person}{Keping Bi},
  \bibinfo{person}{Cheng Luo}, \bibinfo{person}{Jiafeng Guo}, {and}
  \bibinfo{person}{W~Bruce Croft}.} \bibinfo{year}{2018}\natexlab{a}.
\newblock \showarticletitle{Unbiased Learning to Rank with Unbiased Propensity
  Estimation}.
\newblock \bibinfo{journal}{\emph{SIGIR}} (\bibinfo{year}{2018}).
\newblock


\bibitem[\protect\citeauthoryear{Ai, Mao, Liu, and Croft}{Ai
  et~al\mbox{.}}{2018b}]%
        {ai2018unbiasedb}
\bibfield{author}{\bibinfo{person}{Qingyao Ai}, \bibinfo{person}{Jiaxin Mao},
  \bibinfo{person}{Yiqun Liu}, {and} \bibinfo{person}{W~Bruce Croft}.}
  \bibinfo{year}{2018}\natexlab{b}.
\newblock \showarticletitle{Unbiased learning to rank: Theory and practice}. In
  \bibinfo{booktitle}{\emph{CIKM}}.
\newblock


\bibitem[\protect\citeauthoryear{Alaa and van~der Schaar}{Alaa and van~der
  Schaar}{2017}]%
        {alaa2017deep}
\bibfield{author}{\bibinfo{person}{Ahmed~M Alaa} {and} \bibinfo{person}{Mihaela
  van~der Schaar}.} \bibinfo{year}{2017}\natexlab{}.
\newblock \showarticletitle{Deep multi-task gaussian processes for survival
  analysis with competing risks}. In \bibinfo{booktitle}{\emph{NeurIPS}}.
\newblock


\bibitem[\protect\citeauthoryear{Andersen, Borgan, Gill, and Keiding}{Andersen
  et~al\mbox{.}}{2012}]%
        {andersen2012statistical}
\bibfield{author}{\bibinfo{person}{Per~K Andersen}, \bibinfo{person}{Ornulf
  Borgan}, \bibinfo{person}{Richard~D Gill}, {and} \bibinfo{person}{Niels
  Keiding}.} \bibinfo{year}{2012}\natexlab{}.
\newblock \bibinfo{booktitle}{\emph{Statistical models based on counting
  processes}}.
\newblock


\bibitem[\protect\citeauthoryear{Chapelle and Chang}{Chapelle and
  Chang}{2011}]%
        {chapelle2011yahoo}
\bibfield{author}{\bibinfo{person}{Olivier Chapelle} {and} \bibinfo{person}{Yi
  Chang}.} \bibinfo{year}{2011}\natexlab{}.
\newblock \showarticletitle{Yahoo! learning to rank challenge overview}. In
  \bibinfo{booktitle}{\emph{Proceedings of the Learning to Rank Challenge}}.
\newblock


\bibitem[\protect\citeauthoryear{Chapelle and Zhang}{Chapelle and
  Zhang}{2009}]%
        {chapelle2009dynamic}
\bibfield{author}{\bibinfo{person}{Olivier Chapelle} {and} \bibinfo{person}{Ya
  Zhang}.} \bibinfo{year}{2009}\natexlab{}.
\newblock \showarticletitle{A dynamic bayesian network click model for web
  search ranking}. In \bibinfo{booktitle}{\emph{WWW}}.
\newblock


\bibitem[\protect\citeauthoryear{Cox}{Cox}{1972}]%
        {cox1972regression}
\bibfield{author}{\bibinfo{person}{David~R Cox}.}
  \bibinfo{year}{1972}\natexlab{}.
\newblock \showarticletitle{Regression models and life-tables}.
\newblock \bibinfo{journal}{\emph{Journal of the Royal Statistical Society:
  Series B (Methodological)}} (\bibinfo{year}{1972}).
\newblock


\bibitem[\protect\citeauthoryear{Craswell, Zoeter, Taylor, and Ramsey}{Craswell
  et~al\mbox{.}}{2008}]%
        {craswell2008experimental}
\bibfield{author}{\bibinfo{person}{Nick Craswell}, \bibinfo{person}{Onno
  Zoeter}, \bibinfo{person}{Michael Taylor}, {and} \bibinfo{person}{Bill
  Ramsey}.} \bibinfo{year}{2008}\natexlab{}.
\newblock \showarticletitle{An experimental comparison of click position-bias
  models}. In \bibinfo{booktitle}{\emph{WSDM}}.
\newblock


\bibitem[\protect\citeauthoryear{Dupret and Piwowarski}{Dupret and
  Piwowarski}{2008}]%
        {dupret2008user}
\bibfield{author}{\bibinfo{person}{Georges~E Dupret} {and}
  \bibinfo{person}{Benjamin Piwowarski}.} \bibinfo{year}{2008}\natexlab{}.
\newblock \showarticletitle{A user browsing model to predict search engine
  click data from past observations.}. In \bibinfo{booktitle}{\emph{SIGIR}}.
\newblock


\bibitem[\protect\citeauthoryear{Fang, Guo, Zhang, and Shu}{Fang
  et~al\mbox{.}}{2019}]%
        {fang2019deep}
\bibfield{author}{\bibinfo{person}{Hui Fang}, \bibinfo{person}{Guibing Guo},
  \bibinfo{person}{Danning Zhang}, {and} \bibinfo{person}{Yiheng Shu}.}
  \bibinfo{year}{2019}\natexlab{}.
\newblock \showarticletitle{Deep Learning-Based Sequential Recommender Systems:
  Concepts, Algorithms, and Evaluations}. In
  \bibinfo{booktitle}{\emph{International Conference on Web Engineering}}.
\newblock


\bibitem[\protect\citeauthoryear{Fang, Agarwal, and Joachims}{Fang
  et~al\mbox{.}}{2018}]%
        {fang2018intervention}
\bibfield{author}{\bibinfo{person}{Zhichong Fang}, \bibinfo{person}{Aman
  Agarwal}, {and} \bibinfo{person}{Thorsten Joachims}.}
  \bibinfo{year}{2018}\natexlab{}.
\newblock \showarticletitle{Intervention harvesting for context-dependent
  examination-bias estimation}.
\newblock \bibinfo{journal}{\emph{SIGIR}} (\bibinfo{year}{2018}).
\newblock


\bibitem[\protect\citeauthoryear{Gordon and Olshen}{Gordon and Olshen}{1985}]%
        {gordon1985tree}
\bibfield{author}{\bibinfo{person}{Louis Gordon} {and}
  \bibinfo{person}{Richard~A Olshen}.} \bibinfo{year}{1985}\natexlab{}.
\newblock \showarticletitle{Tree-structured survival analysis.}
\newblock \bibinfo{journal}{\emph{Cancer treatment reports}}
  (\bibinfo{year}{1985}).
\newblock


\bibitem[\protect\citeauthoryear{Guo, Liu, Kannan, Minka, Taylor, Wang, and
  Faloutsos}{Guo et~al\mbox{.}}{2009b}]%
        {guo2009click}
\bibfield{author}{\bibinfo{person}{Fan Guo}, \bibinfo{person}{Chao Liu},
  \bibinfo{person}{Anitha Kannan}, \bibinfo{person}{Tom Minka},
  \bibinfo{person}{Michael Taylor}, \bibinfo{person}{Yi-Min Wang}, {and}
  \bibinfo{person}{Christos Faloutsos}.} \bibinfo{year}{2009}\natexlab{b}.
\newblock \showarticletitle{Click chain model in web search}. In
  \bibinfo{booktitle}{\emph{WWW}}.
\newblock


\bibitem[\protect\citeauthoryear{Guo, Liu, and Wang}{Guo
  et~al\mbox{.}}{2009a}]%
        {guo2009efficient}
\bibfield{author}{\bibinfo{person}{Fan Guo}, \bibinfo{person}{Chao Liu}, {and}
  \bibinfo{person}{Yi~Min Wang}.} \bibinfo{year}{2009}\natexlab{a}.
\newblock \showarticletitle{Efficient multiple-click models in web search}. In
  \bibinfo{booktitle}{\emph{WSDM}}.
\newblock


\bibitem[\protect\citeauthoryear{Hochreiter and Schmidhuber}{Hochreiter and
  Schmidhuber}{1997}]%
        {hochreiter1997long}
\bibfield{author}{\bibinfo{person}{Sepp Hochreiter} {and}
  \bibinfo{person}{J{\"u}rgen Schmidhuber}.} \bibinfo{year}{1997}\natexlab{}.
\newblock \showarticletitle{Long short-term memory}.
\newblock \bibinfo{journal}{\emph{Neural computation}} (\bibinfo{year}{1997}).
\newblock


\bibitem[\protect\citeauthoryear{Hu, Wang, Peng, and Li}{Hu
  et~al\mbox{.}}{2019}]%
        {hu2019unbiased}
\bibfield{author}{\bibinfo{person}{Ziniu Hu}, \bibinfo{person}{Yang Wang},
  \bibinfo{person}{Qu Peng}, {and} \bibinfo{person}{Hang Li}.}
  \bibinfo{year}{2019}\natexlab{}.
\newblock \showarticletitle{Unbiased LambdaMART: An Unbiased Pairwise
  Learning-to-Rank Algorithm}. In \bibinfo{booktitle}{\emph{WWW}}.
\newblock


\bibitem[\protect\citeauthoryear{Jagerman, Oosterhuis, and de~Rijke}{Jagerman
  et~al\mbox{.}}{2019}]%
        {jagerman2019model}
\bibfield{author}{\bibinfo{person}{Rolf Jagerman}, \bibinfo{person}{Harrie
  Oosterhuis}, {and} \bibinfo{person}{Maarten de Rijke}.}
  \bibinfo{year}{2019}\natexlab{}.
\newblock \showarticletitle{To Model or to Intervene: A Comparison of
  Counterfactual and Online Learning to Rank from User Interactions}.
\newblock  (\bibinfo{year}{2019}).
\newblock


\bibitem[\protect\citeauthoryear{Jing and Smola}{Jing and Smola}{2017}]%
        {jing2017neural}
\bibfield{author}{\bibinfo{person}{How Jing} {and} \bibinfo{person}{Alexander~J
  Smola}.} \bibinfo{year}{2017}\natexlab{}.
\newblock \showarticletitle{Neural survival recommender}. In
  \bibinfo{booktitle}{\emph{WSDM}}.
\newblock


\bibitem[\protect\citeauthoryear{Joachims, Granka, Pan, Hembrooke, and
  Gay}{Joachims et~al\mbox{.}}{2005}]%
        {joachims2005accurately}
\bibfield{author}{\bibinfo{person}{Thorsten Joachims}, \bibinfo{person}{Laura~A
  Granka}, \bibinfo{person}{Bing Pan}, \bibinfo{person}{Helene Hembrooke},
  {and} \bibinfo{person}{Geri Gay}.} \bibinfo{year}{2005}\natexlab{}.
\newblock \showarticletitle{Accurately interpreting clickthrough data as
  implicit feedback}. In \bibinfo{booktitle}{\emph{SIGIR}}.
\newblock


\bibitem[\protect\citeauthoryear{Joachims, Swaminathan, and Schnabel}{Joachims
  et~al\mbox{.}}{2017}]%
        {joachims2017unbiased}
\bibfield{author}{\bibinfo{person}{Thorsten Joachims}, \bibinfo{person}{Adith
  Swaminathan}, {and} \bibinfo{person}{Tobias Schnabel}.}
  \bibinfo{year}{2017}\natexlab{}.
\newblock \showarticletitle{Unbiased learning-to-rank with biased feedback}. In
  \bibinfo{booktitle}{\emph{WSDM}}.
\newblock


\bibitem[\protect\citeauthoryear{Kaplan and Meier}{Kaplan and Meier}{1958}]%
        {kaplan1958nonparametric}
\bibfield{author}{\bibinfo{person}{Edward~L Kaplan} {and} \bibinfo{person}{Paul
  Meier}.} \bibinfo{year}{1958}\natexlab{}.
\newblock \showarticletitle{Nonparametric estimation from incomplete
  observations}.
\newblock \bibinfo{journal}{\emph{Journal of the American statistical
  association}} (\bibinfo{year}{1958}).
\newblock


\bibitem[\protect\citeauthoryear{Katzman, Shaham, Cloninger, Bates, Jiang, and
  Kluger}{Katzman et~al\mbox{.}}{2018}]%
        {katzman2018deepsurv}
\bibfield{author}{\bibinfo{person}{Jared~L Katzman}, \bibinfo{person}{Uri
  Shaham}, \bibinfo{person}{Alexander Cloninger}, \bibinfo{person}{Jonathan
  Bates}, \bibinfo{person}{Tingting Jiang}, {and} \bibinfo{person}{Yuval
  Kluger}.} \bibinfo{year}{2018}\natexlab{}.
\newblock \showarticletitle{DeepSurv: personalized treatment recommender system
  using a Cox proportional hazards deep neural network}.
\newblock \bibinfo{journal}{\emph{BMC medical research methodology}}
  (\bibinfo{year}{2018}).
\newblock


\bibitem[\protect\citeauthoryear{Ke, Meng, Finley, Wang, Chen, Ma, Ye, and
  Liu}{Ke et~al\mbox{.}}{2017}]%
        {ke2017lightgbm}
\bibfield{author}{\bibinfo{person}{Guolin Ke}, \bibinfo{person}{Qi Meng},
  \bibinfo{person}{Thomas Finley}, \bibinfo{person}{Taifeng Wang},
  \bibinfo{person}{Wei Chen}, \bibinfo{person}{Weidong Ma},
  \bibinfo{person}{Qiwei Ye}, {and} \bibinfo{person}{Tie-Yan Liu}.}
  \bibinfo{year}{2017}\natexlab{}.
\newblock \showarticletitle{Lightgbm: A highly efficient gradient boosting
  decision tree}. In \bibinfo{booktitle}{\emph{NeurIPS}}.
\newblock


\bibitem[\protect\citeauthoryear{Khan and Zubek}{Khan and Zubek}{2008}]%
        {khan2008support}
\bibfield{author}{\bibinfo{person}{Faisal~M Khan} {and}
  \bibinfo{person}{Valentina~Bayer Zubek}.} \bibinfo{year}{2008}\natexlab{}.
\newblock \showarticletitle{Support vector regression for censored data (SVRc):
  a novel tool for survival analysis}. In \bibinfo{booktitle}{\emph{ICDM}}.
\newblock


\bibitem[\protect\citeauthoryear{Lee and Wang}{Lee and Wang}{2003}]%
        {lee2003statistical}
\bibfield{author}{\bibinfo{person}{Elisa~T Lee} {and} \bibinfo{person}{John
  Wang}.} \bibinfo{year}{2003}\natexlab{}.
\newblock \bibinfo{booktitle}{\emph{Statistical methods for survival data
  analysis}}. Vol.~\bibinfo{volume}{476}.
\newblock \bibinfo{publisher}{John Wiley \& Sons}.
\newblock


\bibitem[\protect\citeauthoryear{Li, Huffman, and Tokuda}{Li
  et~al\mbox{.}}{2009}]%
        {li2009good}
\bibfield{author}{\bibinfo{person}{Jane Li}, \bibinfo{person}{Scott Huffman},
  {and} \bibinfo{person}{Akihito Tokuda}.} \bibinfo{year}{2009}\natexlab{}.
\newblock \showarticletitle{Good abandonment in mobile and PC internet search}.
  In \bibinfo{booktitle}{\emph{SIGIR}}.
\newblock


\bibitem[\protect\citeauthoryear{Li, Wang, Ye, and Reddy}{Li
  et~al\mbox{.}}{2016}]%
        {li2016multi}
\bibfield{author}{\bibinfo{person}{Yan Li}, \bibinfo{person}{Jie Wang},
  \bibinfo{person}{Jieping Ye}, {and} \bibinfo{person}{Chandan~K Reddy}.}
  \bibinfo{year}{2016}\natexlab{}.
\newblock \showarticletitle{A multi-task learning formulation for survival
  analysis}. In \bibinfo{booktitle}{\emph{KDD}}.
\newblock


\bibitem[\protect\citeauthoryear{Liu et~al\mbox{.}}{Liu et~al\mbox{.}}{2009}]%
        {liu2009learning}
\bibfield{author}{\bibinfo{person}{Tie-Yan Liu} {et~al\mbox{.}}}
  \bibinfo{year}{2009}\natexlab{}.
\newblock \showarticletitle{Learning to rank for information retrieval}.
\newblock \bibinfo{journal}{\emph{Foundations and Trends{\textregistered} in
  Information Retrieval}} (\bibinfo{year}{2009}).
\newblock


\bibitem[\protect\citeauthoryear{Ranganath, Perotte, Elhadad, and
  Blei}{Ranganath et~al\mbox{.}}{2016}]%
        {ranganath2016deep}
\bibfield{author}{\bibinfo{person}{Rajesh Ranganath}, \bibinfo{person}{Adler
  Perotte}, \bibinfo{person}{No{\'e}mie Elhadad}, {and} \bibinfo{person}{David
  Blei}.} \bibinfo{year}{2016}\natexlab{}.
\newblock \showarticletitle{Deep survival analysis}.
\newblock \bibinfo{journal}{\emph{arXiv}} (\bibinfo{year}{2016}).
\newblock


\bibitem[\protect\citeauthoryear{Ren, Fang, Zhang, Liu, Li, Zhang, Yu, and
  Wang}{Ren et~al\mbox{.}}{2018}]%
        {ren2018learning}
\bibfield{author}{\bibinfo{person}{Kan Ren}, \bibinfo{person}{Yuchen Fang},
  \bibinfo{person}{Weinan Zhang}, \bibinfo{person}{Shuhao Liu},
  \bibinfo{person}{Jiajun Li}, \bibinfo{person}{Ya Zhang},
  \bibinfo{person}{Yong Yu}, {and} \bibinfo{person}{Jun Wang}.}
  \bibinfo{year}{2018}\natexlab{}.
\newblock \showarticletitle{Learning multi-touch conversion attribution with
  dual-attention mechanisms for online advertising}. In
  \bibinfo{booktitle}{\emph{CIKM}}.
\newblock


\bibitem[\protect\citeauthoryear{Ren, Qin, Zheng, Yang, Zhang, Qiu, and Yu}{Ren
  et~al\mbox{.}}{2019a}]%
        {ren2019deepa}
\bibfield{author}{\bibinfo{person}{Kan Ren}, \bibinfo{person}{Jiarui Qin},
  \bibinfo{person}{Lei Zheng}, \bibinfo{person}{Zhengyu Yang},
  \bibinfo{person}{Weinan Zhang}, \bibinfo{person}{Lin Qiu}, {and}
  \bibinfo{person}{Yong Yu}.} \bibinfo{year}{2019}\natexlab{a}.
\newblock \showarticletitle{Deep recurrent survival analysis}. In
  \bibinfo{booktitle}{\emph{AAAI}}.
\newblock


\bibitem[\protect\citeauthoryear{Ren, Qin, Zheng, Zhang, and Yu}{Ren
  et~al\mbox{.}}{2019b}]%
        {ren2019deepb}
\bibfield{author}{\bibinfo{person}{Kan Ren}, \bibinfo{person}{Jiarui Qin},
  \bibinfo{person}{Lei Zheng}, \bibinfo{person}{Weinan Zhang}, {and}
  \bibinfo{person}{Yong Yu}.} \bibinfo{year}{2019}\natexlab{b}.
\newblock \showarticletitle{Deep Landscape Forecasting for Real-time Bidding
  Advertising}.
\newblock \bibinfo{journal}{\emph{KDD}} (\bibinfo{year}{2019}).
\newblock


\bibitem[\protect\citeauthoryear{Richardson, Dominowska, and Ragno}{Richardson
  et~al\mbox{.}}{2007}]%
        {richardson2007predicting}
\bibfield{author}{\bibinfo{person}{Matthew Richardson}, \bibinfo{person}{Ewa
  Dominowska}, {and} \bibinfo{person}{Robert Ragno}.}
  \bibinfo{year}{2007}\natexlab{}.
\newblock \showarticletitle{Predicting clicks: estimating the click-through
  rate for new ads}. In \bibinfo{booktitle}{\emph{WWW}}.
\newblock


\bibitem[\protect\citeauthoryear{Rosenbaum and Rubin}{Rosenbaum and
  Rubin}{1983}]%
        {rosenbaum1983central}
\bibfield{author}{\bibinfo{person}{Paul~R Rosenbaum} {and}
  \bibinfo{person}{Donald~B Rubin}.} \bibinfo{year}{1983}\natexlab{}.
\newblock \showarticletitle{The central role of the propensity score in
  observational studies for causal effects}.
\newblock \bibinfo{journal}{\emph{Biometrika}} (\bibinfo{year}{1983}).
\newblock


\bibitem[\protect\citeauthoryear{Song, Shi, White, and Awadallah}{Song
  et~al\mbox{.}}{2014}]%
        {song2014context}
\bibfield{author}{\bibinfo{person}{Yang Song}, \bibinfo{person}{Xiaolin Shi},
  \bibinfo{person}{Ryen White}, {and} \bibinfo{person}{Ahmed~Hassan
  Awadallah}.} \bibinfo{year}{2014}\natexlab{}.
\newblock \showarticletitle{Context-aware web search abandonment prediction}.
  In \bibinfo{booktitle}{\emph{SIGIR}}.
\newblock


\bibitem[\protect\citeauthoryear{Tibshirani}{Tibshirani}{1997}]%
        {tibshirani1997lasso}
\bibfield{author}{\bibinfo{person}{Robert Tibshirani}.}
  \bibinfo{year}{1997}\natexlab{}.
\newblock \showarticletitle{The lasso method for variable selection in the Cox
  model}.
\newblock \bibinfo{journal}{\emph{Statistics in medicine}}
  (\bibinfo{year}{1997}).
\newblock


\bibitem[\protect\citeauthoryear{Wang, Liu, Wang, Zhou, Nie, and Ma}{Wang
  et~al\mbox{.}}{2015}]%
        {wang2015incorporating}
\bibfield{author}{\bibinfo{person}{Chao Wang}, \bibinfo{person}{Yiqun Liu},
  \bibinfo{person}{Meng Wang}, \bibinfo{person}{Ke Zhou},
  \bibinfo{person}{Jian-yun Nie}, {and} \bibinfo{person}{Shaoping Ma}.}
  \bibinfo{year}{2015}\natexlab{}.
\newblock \showarticletitle{Incorporating non-sequential behavior into click
  models}. In \bibinfo{booktitle}{\emph{SIGIR}}.
\newblock


\bibitem[\protect\citeauthoryear{Wang, Zhai, Dong, and Chang}{Wang
  et~al\mbox{.}}{2013}]%
        {wang2013content}
\bibfield{author}{\bibinfo{person}{Hongning Wang}, \bibinfo{person}{ChengXiang
  Zhai}, \bibinfo{person}{Anlei Dong}, {and} \bibinfo{person}{Yi Chang}.}
  \bibinfo{year}{2013}\natexlab{}.
\newblock \showarticletitle{Content-aware click modeling}. In
  \bibinfo{booktitle}{\emph{WWW}}.
\newblock


\bibitem[\protect\citeauthoryear{Wang, De~Vries, and Reinders}{Wang
  et~al\mbox{.}}{2006}]%
        {wang2006user}
\bibfield{author}{\bibinfo{person}{Jun Wang}, \bibinfo{person}{Arjen~P
  De~Vries}, {and} \bibinfo{person}{Marcel~JT Reinders}.}
  \bibinfo{year}{2006}\natexlab{}.
\newblock \showarticletitle{A user-item relevance model for log-based
  collaborative filtering}. In \bibinfo{booktitle}{\emph{ECIR}}.
\newblock


\bibitem[\protect\citeauthoryear{Wang, Li, and Reddy}{Wang
  et~al\mbox{.}}{2019}]%
        {wang2019machine}
\bibfield{author}{\bibinfo{person}{Ping Wang}, \bibinfo{person}{Yan Li}, {and}
  \bibinfo{person}{Chandan~K Reddy}.} \bibinfo{year}{2019}\natexlab{}.
\newblock \showarticletitle{Machine learning for survival analysis: A survey}.
\newblock \bibinfo{journal}{\emph{CSUR}} (\bibinfo{year}{2019}).
\newblock


\bibitem[\protect\citeauthoryear{Wang, Bendersky, Metzler, and Najork}{Wang
  et~al\mbox{.}}{2016}]%
        {wang2016learning}
\bibfield{author}{\bibinfo{person}{Xuanhui Wang}, \bibinfo{person}{Michael
  Bendersky}, \bibinfo{person}{Donald Metzler}, {and} \bibinfo{person}{Marc
  Najork}.} \bibinfo{year}{2016}\natexlab{}.
\newblock \showarticletitle{Learning to rank with selection bias in personal
  search}. In \bibinfo{booktitle}{\emph{SIGIR}}.
\newblock


\bibitem[\protect\citeauthoryear{Wang, Golbandi, Bendersky, Metzler, and
  Najork}{Wang et~al\mbox{.}}{2018}]%
        {wang2018position}
\bibfield{author}{\bibinfo{person}{Xuanhui Wang}, \bibinfo{person}{Nadav
  Golbandi}, \bibinfo{person}{Michael Bendersky}, \bibinfo{person}{Donald
  Metzler}, {and} \bibinfo{person}{Marc Najork}.}
  \bibinfo{year}{2018}\natexlab{}.
\newblock \showarticletitle{Position bias estimation for unbiased learning to
  rank in personal search}. In \bibinfo{booktitle}{\emph{WSDM}}.
\newblock


\bibitem[\protect\citeauthoryear{Yang, Dai, Yang, Carbonell, Salakhutdinov, and
  Le}{Yang et~al\mbox{.}}{2019}]%
        {yang2019xlnet}
\bibfield{author}{\bibinfo{person}{Zhilin Yang}, \bibinfo{person}{Zihang Dai},
  \bibinfo{person}{Yiming Yang}, \bibinfo{person}{Jaime Carbonell},
  \bibinfo{person}{Ruslan Salakhutdinov}, {and} \bibinfo{person}{Quoc~V Le}.}
  \bibinfo{year}{2019}\natexlab{}.
\newblock \showarticletitle{XLNet: Generalized Autoregressive Pretraining for
  Language Understanding}.
\newblock \bibinfo{journal}{\emph{arXiv}} (\bibinfo{year}{2019}).
\newblock


\bibitem[\protect\citeauthoryear{Zhang, Wu, Che, Lin, Memisevic, Salakhutdinov,
  and Bengio}{Zhang et~al\mbox{.}}{2016a}]%
        {zhang2016architectural}
\bibfield{author}{\bibinfo{person}{Saizheng Zhang}, \bibinfo{person}{Yuhuai
  Wu}, \bibinfo{person}{Tong Che}, \bibinfo{person}{Zhouhan Lin},
  \bibinfo{person}{Roland Memisevic}, \bibinfo{person}{Ruslan~R Salakhutdinov},
  {and} \bibinfo{person}{Yoshua Bengio}.} \bibinfo{year}{2016}\natexlab{a}.
\newblock \showarticletitle{Architectural complexity measures of recurrent
  neural networks}. In \bibinfo{booktitle}{\emph{NeurIPS}}.
\newblock


\bibitem[\protect\citeauthoryear{Zhang, Zhou, Wang, and Xu}{Zhang
  et~al\mbox{.}}{2016b}]%
        {zhang2016bidaware}
\bibfield{author}{\bibinfo{person}{Weinan Zhang}, \bibinfo{person}{Tianxiong
  Zhou}, \bibinfo{person}{Jun Wang}, {and} \bibinfo{person}{Jian Xu}.}
  \bibinfo{year}{2016}\natexlab{b}.
\newblock \showarticletitle{Bid-aware Gradient Descent for Unbiased Learning
  with Censored Data in Display Advertising}. In
  \bibinfo{booktitle}{\emph{KDD}}.
\newblock


\bibitem[\protect\citeauthoryear{Zhu, Li, Zhang, Li, He, Li, and Gai}{Zhu
  et~al\mbox{.}}{2018}]%
        {zhu2018learning}
\bibfield{author}{\bibinfo{person}{Han Zhu}, \bibinfo{person}{Xiang Li},
  \bibinfo{person}{Pengye Zhang}, \bibinfo{person}{Guozheng Li},
  \bibinfo{person}{Jie He}, \bibinfo{person}{Han Li}, {and}
  \bibinfo{person}{Kun Gai}.} \bibinfo{year}{2018}\natexlab{}.
\newblock \showarticletitle{Learning Tree-based Deep Model for Recommender
  Systems}. In \bibinfo{booktitle}{\emph{KDD}}.
\newblock


\end{thebibliography}

\end{document}